\documentclass[a4paper,12pt]{article}%
\usepackage{amsmath}
\usepackage{amsfonts}
\usepackage{mathrsfs}
\usepackage{amssymb}
\usepackage{amsthm}
\usepackage{graphics}
\usepackage{mathtools}
\usepackage{bm}
\usepackage{centernot}
\usepackage{oubraces}
\usepackage{cancel}
\usepackage{enumitem}
\usepackage{mathrsfs}
\usepackage{caption}
\usepackage{subcaption}
\usepackage{listings}
\usepackage{color}
\usepackage{colortbl}
\usepackage{tikz}
\usetikzlibrary{shapes.geometric}
\usetikzlibrary{arrows,arrows.meta}
\usepackage{eso-pic}
\everymath{\displaystyle}
\tikzstyle{arrow}=[arrows={-Stealth[scale=0.7]}]
\tikzstyle{shadow}=[preaction={fill=black,opacity=.5,transform canvas={xshift=0.5mm,yshift=-0.5mm},shading=radial,shading angle=20},fill=red]

\tikzset{
    treenode/.style = {shape=rectangle, rounded corners,
        draw, align=center,
        top color=white, bottom color=blue!20},
    root/.style     = {treenode, font=\footnotesize, bottom color=blue!20},
    env/.style      = {treenode, font=\ttfamily\footnotesize},
}
\usepackage{indentfirst}
\usepackage[margin=1in]{geometry}
\usepackage{times}
\usepackage{xcolor}
\usepackage{soul}
\setstcolor{red}
\usepackage{natbib}
\usepackage{rotating}
\usepackage{hyperref}%
\providecommand{\U}[1]{\protect\rule{.1in}{.1in}}

\usepackage[normalem]{ulem}

\hypersetup{bookmarksopen=true,citecolor=blue,
    linkcolor=blue,colorlinks,bookmarksopenlevel=2,pdfstartview=Fit,
    pdftitle={Equity in Obviously Strategy-proof Exchange}, pdfauthor={Peng Liu and Huaxia Zeng}}

\renewcommand{\cite}{\citet}
\renewcommand{\leq}{\leqslant}
\renewcommand{\geq}{\geqslant}

\arrayrulecolor{cyan}
\definecolor{OliveGreen}{cmyk}{0.64,0,0.95,0.40}

\makeatother

\newtheorem{theorem}{Theorem}
\newtheorem{definition}{Definition}

\newtheorem{lemma}{Lemma}
\newtheorem{proposition}{Proposition}
\newtheorem{example}{Example}

\newtheorem{observation}{Observation}

\makeatletter
\DeclareFontFamily{U}{MnSymbolA}{}
\DeclareFontShape{U}{MnSymbolA}{m}{n}{
    <-6>  MnSymbolA5
   <6-7>  MnSymbolA6
   <7-8>  MnSymbolA7
   <8-9>  MnSymbolA8
   <9-10> MnSymbolA9
  <10-12> MnSymbolA10
  <12->   MnSymbolA12}{}
\DeclareFontShape{U}{MnSymbolA}{b}{n}{
    <-6>  MnSymbolA-Bold5
   <6-7>  MnSymbolA-Bold6
   <7-8>  MnSymbolA-Bold7
   <8-9>  MnSymbolA-Bold8
   <9-10> MnSymbolA-Bold9
  <10-12> MnSymbolA-Bold10
  <12->   MnSymbolA-Bold12}{}
\DeclareSymbolFont{MnSyA}{U}{MnSymbolA}{m}{n}
\SetSymbolFont{MnSyA}{bold}{U}{MnSymbolA}{b}{n}

\DeclareRobustCommand{\overleftharpoon}{\mathpalette{\overarrow@\leftharpoonfill@}}
\DeclareRobustCommand{\overrightharpoon}{\mathpalette{\overarrow@\rightharpoonfill@}}
\def\leftharpoonfill@{\arrowfill@\leftharpoondown\mn@relbar\mn@relbar}
\def\rightharpoonfill@{\arrowfill@\mn@relbar\mn@relbar\rightharpoonup}

\DeclareMathSymbol{\leftharpoondown}{\mathrel}{MnSyA}{'112}
\DeclareMathSymbol{\rightharpoonup}{\mathrel}{MnSyA}{'100}
\DeclareMathSymbol{\mn@relbar}{\mathrel}{MnSyA}{'320}
\makeatother

\makeatletter
\def\overrightharpoonfill@{\arrowfill@\relbar\relbar\rightharpoonup}
\DeclareRobustCommand{\overrightharpoon}{\mathpalette{\overarrow@\overrightharpoonfill@}}

\def\downrightharpoonfill@{\arrowfill@\relbar\relbar\rightharpoondown}
\DeclareRobustCommand{\downrightharpoon}{\mathpalette{\raise0.2em\underarrow@\downrightharpoonfill@}}
\makeatother

\begin{document}
    
    \title{Equity in Obviously Strategy-Proof Exchange\thanks{We are grateful to William Thomson and Jingyi Xue for their insightful comments. We also thank the participants of the 10th PKU-NUS Annual International Conference on Quantitative Finance and Economics at Peking University HSBC Business School, 
    the 2026 CSW Meeting of the Econometric Society at New York University Abu Dhabi,
    the 2025 Shanghai Microeconomics Workshop at SUFE and the 14th Conference on Economic Design at the University of Essex for their helpful suggestions. Huaxia Zeng acknowledges that his work was supported by 
    the National Natural Science Foundation of China (No.~72573105),
    the National Natural Science Foundation of China (No.~72394391), and the Program for Professor of Special Appointment (Eastern Scholar) at Shanghai Institutions of Higher Learning (No.~2019140015).}}
    \author{Peng Liu\thanks{School of Economics and Management, East China Normal University, Shanghai 200062, China. E-mail: pengliu0618@gmail.com.} ~~and~ Huaxia Zeng\thanks{School of Economics, Shanghai University of Finance and Economics, Shanghai 200433, China. E-mail: huaxia.zeng.2011@gmail.com.}}
    \date{\today}
    \maketitle
    
    \vspace{-1em}
    \begin{abstract}
    \small
        \noindent
        We propose a novel hierarchy of merit-based fairness axioms for designing exchange rules: Equity under Acclaims (EA) $\Rightarrow$ Equity under Bipartite Acclaims (EBA) $\Rightarrow$ Weak Equity under Bipartite Acclaims (WEBA) $\Rightarrow$ Equity under Unanimous Acclaims (EUA).
        With this hierarchy, we identify two unambiguous costs of strengthening strategy-proofness to obvious strategy-proofness. First, the applicability of obviously strategy-proof rules is narrowed: under a mild domain richness condition, an obviously strategy-proof rule satisfying EUA, the weakest fairness requirement in our hierarchy, exists if and only if preferences are single-peaked. By contrast, TTC, a strategy-proof rule under unrestricted preferences, satisfies our strongest fairness requirement, EA. Second, the level of attainable fairness is substantially reduced: we establish that no obviously strategy-proof rule satisfies EBA, whereas WEBA is attainable by a new obviously strategy-proof rule, Designator.
        
        
    \end{abstract}

\newpage

{\centering\section{Introduction}\label{sec:Introduction}}

The market design literature increasingly values mechanisms that are simple enough for real participants to understand and use correctly. Obvious strategy-proofness, introduced by \cite{L2017}, captures this ideal: an agent needs only to verify that the worst outcome under one action is no worse than the best outcome under any alternative, without engaging in contingent reasoning. This cognitive simplicity, however, may come at a cost. In this paper, we investigate the costs of strengthening the conventional incentive requirement of strategy-proofness to obvious strategy-proofness in the classic exchange problem.

The strategic exchange model for indivisible goods, formalized by \cite{SS1974}, is a cornerstone of market design.
The Top Trading Cycles (TTC) rule, introduced in the same paper and attributed to David Gale, has been the benchmark mechanism for over five decades.
Its canonical status rests on a landmark characterization by \cite{M1994}: on the domain of strict and otherwise unrestricted preferences, TTC is the unique rule that simultaneously satisfies efficiency, individual rationality, and strategy-proofness.
This trifecta of normative and incentive properties has made TTC the reference point not only for exchange, but also for kidney exchange \citep{RSU2004}, house allocation with existing tenants \citep{AS1999}, school choice \citep{AS2003}, and general object allocation \citep{P2000}.

Yet one aspect of TTC's appeal has not been formally analyzed: its implicit fairness. The TTC algorithm rewards agents whose endowments generate the greatest value for others, reflecting a merit-based principle of distributive justice that can be traced back to Aristotle's \textit{Nicomachean Ethics} [\href{https://classics.mit.edu/Aristotle/nicomachaen.5.v.html}{\textcolor{blue}{Book V, Chapter 3}}]: ``\textit{... the unjust is unequal, just is equal, as all men suppose it to be, even apart from argument. ... Further, this is plain from the fact that awards should be `according to merit'; for all men agree that what is just in distribution must be according to merit in some sense.}'' Despite this natural connection, no formal fairness criterion has been proposed to explicitly study TTC's fairness performance.\footnote{Unlike the strategic exchange setting studied in this paper, fairness is a central concern in many other allocation problems. For instance, in school choice, the fairness properties of TTC-based matching rules have been extensively studied from the perspective of eliminating justified envy \citep[\emph{e.g.},][]{M2015}. 
For another instance, a fundamental tension between incentive compatibility and natural fairness constraints is established by \cite{TS2010}. They show no strategy-proof matching rule can satisfy respect for pairwise unanimity—the requirement that mutually top-ranked agents must be matched together.
However, those notions of fairness are not applicable to the endowment-exchange problems considered here.}

Although TTC is strategy-proof, \cite{L2017} showed that it is not obviously strategy-proof, while multiple laboratory and field experiments have documented systematic misreporting under TTC \citep{CS2006,GH2018,HK2018}.
Alternatively, \cite{B2019} introduced a new exchange rule: Crawler, and showed that restricting preferences to be single-peaked with respect to an exogenous linear order, it restores the compatibility of efficiency, individual rationality and obvious strategy-proofness.
Subsequent work has expanded the set of obviously strateg-proof exchange rules under single-peaked preferences \citep[\emph{e.g.},][]{T2023}.

These recent developments create two interconnected gaps.
First, obviously strategy-proof rules, like Crawler, abandon the trading-cycle structure that underpins TTC's merit-based fairness, suggesting that strengthening strategy-proofness to obvious strategy-proofness may entail a fairness sacrifice.
Yet, no work has systematically evaluated the fairness performance of obviously strategy-proof exchange rules, making the omission of fairness analysis in exchange even more striking.
Second, single-peakedness is introduced as an exogenous assumption for analyzing obviously strategy-proof exchange rules; no theoretical justification is provided for its imposition.
Our results address both gaps.

We propose a hierarchy of merit-based fairness axioms.
These axioms recognize \textit{acclaimed agents}: active participants whose endowments are the most preferred active objects of at least one other active agent.\footnote{We repeatedly remove agents who most-prefer their endowments, among the remaining ones.
Under the requirement of individual rationality, these agents receive their own objects and hence are inactive for exchange.
The agents who remain are active for exchange, called the active agents, and their endowments are called the active objects.}
Our strongest axiom, Equity under Acclaims (EA), requires each acclaimed agent to receive an object no worse than the endowments of those who acclaim her object.
TTC clearly satisfies EA.
We then focus on the case of exactly two acclaimed agents, where EA reduces to Equity under Bipartite Acclaims (EBA), requiring both acclaimed agents to receive their favorite active objects. 
We further introduce two nested weakenings. Weak Equity under Bipartite Acclaims (WEBA) requires at least one member of the two acclaimed agents to receive her favorite active object, and favors the unanimously acclaimed agent (if exists) - the acclaimed agent whose object is unanimously acclaimed by all other active agents. Last, Equity under Unanimous Acclaims (EUA), the weakest axiom in our hierarchy, requires only that the unanimously acclaimed agent receive her favorite active object.

Our analysis yields three main results, which together uncover two unambiguous costs of strengthening strategy-proofness to obvious strategy-proofness.

First, Theorem~\ref{thm:merit} shows that obvious strategy-proofness, combined with even the weakest axiom in our hierarchy forces a restriction on the domain of admissible preferences.
Specifically, we prove that on a rich domain,\footnote{Richness of a domain refers to the combination of path-connectedness and a diversity condition.
Both are standard in related literature.
For details, see Section~\ref{sec:unanimous}.}
an efficient, individually rational, obviously strategy-proof rule satisfying EUA exists if and only if the domain consists exclusively of single-peaked preferences.
This provides the first theoretical justification for the role of single-peakedness in the literature on obviously strategy-proof mechanism design: the restriction is not an ad hoc assumption, but a necessary condition for reconciling obvious strategy-proofness with any merit-based fairness requirement no weaker than EUA.
Thus, the \textbf{first cost} is a narrowing of the applicability of obviously strategy-proof rules.
In particular, obviously strategy-proof rules with merit-based fairness can only operate in an environment where single-peaked preferences are plausible, whereas TTC is strategy-proof and fair on the unrestricted domain.

Second, Theorem~\ref{thm:impossibility} shows that even under single-peaked preferences, fairness cannot be fully restored.
For economies with at least four agents, no efficient, individually rational, obviously strategy-proof rule satisfies EBA.

Last, we introduce a new exchange rule: \emph{Designator}, which blends the sequential structure of Crawler with the merit-based fairness embodied by TTC.
Theorem~\ref{thm:designator} establishes that Designator satisfies efficiency, individual rationality, obvious strategy-proofness, and WEBA on the domain of single-peaked preferences.
Together with Theorem~\ref{thm:impossibility}, this reveals the \textbf{second cost}---a substantial reduction in fairness: on the domain of single-peaked preferences, WEBA is a fairness upper bound attainable by obviously strategy-proof rules, which however can never match the fairness standard delivered by TTC.

The paper proceeds as follows.
Section~\ref{sec:model} presents the model and preliminaries.
Section~\ref{sec:hierarchy} introduces the fairness axioms.
Section~\ref{sec:unanimous} proves Theorem~\ref{thm:merit} and identifies the first cost.
Section~\ref{sec:upperbound} proves Theorems~\ref{thm:impossibility} and~\ref{thm:designator} and introduces Designator, establishing the second cost.
Section~\ref{sec:conclusion} concludes.
All proofs are in the Appendix.

{\centering\section{Model and Preliminaries}\label{sec:model}}

Let $N=\{1,\dots,n\}$ be a finite set of agents, where $n \geq 3$.
Each agent $i\in N$ owns an object, denoted $o_i$.
The set of objects for exchange is hence $O=\{o_1,\dots,o_n\}$.
An \textbf{allocation}, denoted $m$, is a one-to-one mapping from $N$ to $O$,
where $m(i)$ denotes the object allocated to agent $i$.
By definition, the endowment is also an allocation, denoted $e$, such that $e(i)=o_i$ for all $i \in N$.
Let $\mathcal{M}$ be the set of all allocations.

Each agent $i\in N$ has a strict preference $P_i$ over objects, \textit{i.e.}, a complete, anti-symmetric and transitive binary relation over $O$.
Let $R_i$ be the weak preference, \textit{i.e.}, $o\mathrel{R_i}o'$ if and only if $o\mathrel{P_i}o'$ or $o=o'$.
Let $r_k(P_i)$ be the $k$th ranked object according to $P_i$.
Moreover, given a non-empty subset $O' \subseteq O$, let $\max\nolimits^{P_i}O'$ and $\min\nolimits^{P_i}O'$ denote respectively the most and least preferred objects in $O'$ according to $P_i$.
Let $\mathcal{P}$ be the set of all preferences.
For specific problems, some preferences may not be admissible.
A set of admissible preferences $\mathcal{D}\subseteq\mathcal{P}$ is called a \textbf{preference domain}.
Henceforth, we call $\mathcal{P}$ \textbf{the unrestricted domain}.

An exchange rule, or simply a \textbf{rule}, is a mapping $f:\mathcal{D}^n\rightarrow \mathcal{M}$ that selects for each profile of admissible preferences $P\in \mathcal{D}^n$ an allocation $f(P)\in \mathcal{M}$.
Let $f_i(P)$ denote the object allocated to agent $i$ according to the allocation $f(P)$.
Three desirable properties of rules are as follows.
First, a rule $f:\mathcal{D}^n\rightarrow \mathcal{M}$ is \textbf{efficient} if, for each $P \in \mathcal{D}^n$, the chosen allocation $f(P)$ is Pareto efficient, \textit{i.e.}, there exists no $m' \in \mathcal{M}$ such that $m'(i) \mathrel{R_i} f_i(P)$ for all $i \in N$, and $m'(j) \mathrel{P_j} f_j(P)$ for some $j \in N$.
Second, a rule $f:\mathcal{D}^n\rightarrow \mathcal{M}$ is \textbf{individually rational} if, for each $P \in \mathcal{D}^n$, the chosen allocation $f(P)$ is individually rational, \textit{i.e.}, $f_i(P)\mathrel{R_i} o_i$ for all $i \in N$.
Third, a rule $f: \mathcal{D}^n \rightarrow \mathcal{M}$ is \textbf{strategy-proof} if for all $i \in N$, $P_i, P_i' \in \mathcal{D}$ and $P_{-i} \in \mathcal{D}^{n-1}$, 
$f_i(P_i, P_{-i})\mathrel{R_i} f_i(P_i', P_{-i})$, which implies that in the associated preference revelation game, reporting truthfully is always a weakly dominant strategy for each agent. 

\cite{SS1974} introduced the top trading cycles rule (TTC),
and attributed it to David Gale.
Given an arbitrary preference profile, TTC determines the final allocation through the following algorithm.
Each agent ``points'' to the owner of her favorite object.
There must exist at least one cycle, including self-pointing.
All agents involved in a cycle exchange their objects along the cycle and leave.
If some agents remain, each agent points to the owner of her favorite object among the remaining ones. The procedure proceeds until all agents leave.
TTC is well known to be the unique rule satisfying these properties on the unrestricted domain.\footnote{\cite{M2013} provides alternative characterizations with incentive compatibility notions weaker than strategy-proofness and a consistency requirement that an agent’s final allocation stays unchanged if she only reorders objects ranked below than her allocated object.}

\bigskip
\begin{proposition}[\citealp{M1994}]
    \label{prop:M1994}
    On the unrestricted domain $\mathcal{P}$,  TTC is the unique rule that is efficient, individually rational and strategy-proof.
\end{proposition}

{\centering\subsection{Obvious Strategy-Proofness}\label{subsec:OSP}}

Throughout the paper, we focus on \textit{obviously strategy-proof} exchange rules, which substantially strengthen the incentive properties of strategy-proof rules.
To introduce the notion of obvious strategy-proofness, we first need to define an extensive game form.
An \textbf{extensive game form} is a tuple $\Gamma=\langle N,\mathcal{H},\rho,X\rangle$, 
consisting of 
(1) the set of players $N$,
(2) the set of histories $\mathcal{H}$ that are partially ordered on a semi-lattice $\subseteq$,
where $h_{\emptyset} \equiv \textrm{inf}^{\subseteq} \,\mathcal{H}$ is the root of $\Gamma$, and
$\mathcal{T}\subset \mathcal{H}$ denotes the set of terminal histories, \textit{i.e.}, $h \in \mathcal{T}$ if and only if there exists no $h' \in \mathcal{H}$ such that $h \subset h'$,
(3) the player function $\rho:\mathcal{H}\backslash \mathcal{T} \rightarrow N$ that assigns to each non-terminal history a player, and
(4) the outcome function $X:\mathcal{T}\rightarrow \mathcal{M}$ that determines for each terminal history an allocation,
where $X_i(h)$ represents the object allocated to $i$ by $X(h)$.

For each player $i\in N$, let $\mathcal{H}_i \equiv \{ h\in \mathcal{H}\backslash \mathcal{T}: \rho(h)=i \}$
denote the set of histories where $i$ is called to act.
Given $h_i\in \mathcal{H}_i$, let $\mathcal{A}(h_i)\equiv \{a: (h_i,a) \in \mathcal{H} \}$ denote the set of feasible actions for agent $i$ at $h_i$.
A strategy of player $i$, denoted $s_i$, is a function that chooses at each $i$'s history a feasible action, \textit{i.e.}, $s_i(h_i) \in \mathcal{A}(h_i)$ for all $h_i \in \mathcal{H}_i$.
Let $S_i$ denote the set of all agent $i$'s strategies.
A strategy profile is an $n$-tuple $s =(s_1, \dots, s_n)\in \times_{i\in N} S_i$.
Given a non-terminal history $h$ and a strategy profile $s$,
let $z^{\Gamma}(h, s)$ denote the terminal history that is uniquely reached in $\Gamma$ when the game starts at $h$ and proceeds according to $s$.
In particular, if $h=h_{\emptyset}$, we simply write $z^{\Gamma}(s)$.
An object $o$ is said \textit{feasible} for a player $i$ at a history $h_i \in \mathcal{H}_i$ by following a strategy $s_i$ if there exists $s_{-i} \in S_{-i}$ such that $X_i\big(z^{\Gamma}(h_i, (s_i, s_{-i}))\big)= o$.
Then, let $X_i(h_i, s_i) \equiv \big\{o \in O: o = X_i\big(z^{\Gamma}(h_i, (s_i, s_{-s}))\big)\; \textrm{for some}\; s_{-i} \in S_{-i}\big\}$ be the \textit{feasible set} that collects all feasible objects for agent $i$ at $h_i$ given the strategy $s_i$.
Moreover, let $\alpha(s_i)\equiv\{ h_i\in\mathcal{H}_i: h_i\subseteq z^{\Gamma}(s_i,s_{-i})\text{ for some }s_{-i}\in S_{-i} \}$ collect all \textit{on-path} histories of player $i$ according to the strategy $s_i$.

Now, we are ready to present the notion of obvious dominance.
Given $i \in N$ and two strategies $s_i, s_i' \in S_i$, we compare $s_i$ and $s_i'$ as follows.
We first identify all on-path histories where $s_i$ and $s_i'$ choose different actions:
$\mathcal{E}^{\Gamma}(s_i, s_i') \equiv\{ h_i\in \alpha(s_i)\cap \alpha(s_i'): s_i(h_i)\neq s_i'(h_i) \}$.
In fact,  $\mathcal{E}^{\Gamma}(s_i, s_i')$ contains the \textit{earliest departures} of $s_i$ and $s_i'$.\footnote{Given $h_i \in \mathcal{E}^{\Gamma}(s_i, s_i')$ and $h_i' \in \alpha(s_i)\cap \alpha(s_i')$ such that $h_i' \subset h_i$, 
one can easily observe that $s_i$ and $s_i'$ must choose the same action at $h_i'$.}
Then, the strategy $s_i$ is said to \textbf{obviously dominate} $s_i'$ at a preference $P_i$
if at each $h_i \in \mathcal{E}^{\Gamma}(s_i, s_i')$,
the worst feasible outcome given $s_i$ is no worse than the best feasible outcome given $s_i'$, \textit{i.e.},
$\mathop{\min\nolimits^{P_i}}X_i(h_i, s_i)\mathrel{R_i}
\mathop{\max\nolimits^{P_i}}X_i(h_i, s_i')$.\footnote{Notice that if $s_i(h_i) = s_i'(h_i)$ for all $h_i\in \alpha(s_i)\cap \alpha(s_i')$, then $\mathcal{E}^{\Gamma}(s_i, s_i') = \emptyset$. In this case, $s_i$ and $s_i'$ are effectively identical, \emph{i.e.}, $z^{\Gamma}(s_i, s_{-i}) = z^{\Gamma}(s_i', s_{-i})$ for all $s_{-i} \in S_{-i}$, and hence the obvious dominance relation between the two strategies holds vacuously.}
Accordingly, a strategy $s_i$ is an \textbf{obviously dominant strategy} at $P_i$
if it obviously dominates every other strategy of $S_i$.

Fixing an extensive game form $\Gamma$,
each agent $i$ prepares a \textbf{plan} $\mathcal{S}_i: \mathcal{D} \rightarrow S_i$ that chooses a strategy for each preference in the domain $\mathcal{D}$.
For notational convenience, we henceforth write $\mathcal{S}_i^{P_i} \equiv \mathcal{S}_i(P_i)$ to denote the strategy obtained by following the plan $\mathcal{S}_i$ at $P_i$.
\bigskip

\begin{definition}\label{def:OSP}
    A rule $f:\mathcal{D}^n\rightarrow \mathcal{M}$ is \textbf{obviously strategy-proof}
    if there exist an extensive game form $\Gamma$ and plans $\mathcal{S}_1, \dots, \mathcal{S}_n$ such that the following two conditions are satisfied:
    \begin{itemize}
        \item[\rm (i)] for each $i \in N$ and $P_i \in \mathcal{D}$,
        $\mathcal{S}_i^{P_i}$ is an obviously dominant strategy at $P_i$, and
        
        \item[\rm (ii)] for each $(P_1, \dots, P_n)\in\mathcal{D}^n$, $f(P_1, \dots, P_n)
        = X\big(z^{\Gamma}(\mathcal{S}_1^{P_1}, \dots, \mathcal{S}_n^{P_n})\big)$.
    \end{itemize}
    Correspondingly, we say that $\Gamma$ and $\mathcal{S}_1, \dots, \mathcal{S}_n$ \textbf{OSP-implement} $f$.
\end{definition}

By definition, obvious strategy-proofness implies strategy-proofness.

\cite{L2017} showed that TTC is not obviously strategy-proof on the unrestricted domain.\footnote{Besides strategic exchange, 
obvious strategy-proofness has been applied to many other mechanism design problems, \textit{e.g.}, strategic voting \citep{BG2016,AMN2020}, fair division \citep{AMN2023}, object allocation \citep{MR2022}, random assignment \citep{PT2024}, two-sided matching \citep{AG2018} and auction \citep{R2024}.
A revelation principle for obvious strategy-proofness was provided by \cite{M2020}.
Moreover, the idea of obvious dominance has been generalized to address simplicity of mechanisms \citep{PT2023,L2024}.}
Hence, in conjunction with Proposition~\ref{prop:M1994},
we obtain the following impossibility result.

\begin{proposition}
    \label{prop:ML}
    On the unrestricted domain $\mathcal{P}$, no rule is efficient, individually rational and obviously strategy-proof.
\end{proposition}

{\centering\subsection{Crawler and the Single-Peaked Preferences}}

\citet{B2019} introduced a new exchange rule called Crawler, which is defined with respect to an exogenously given linear order over objects.
For instance, in the housing market introduced by \cite{SS1974}, 
the linear order over houses could be established according to their size.\footnote{For another example, suppose that the houses for exchange are located along a main street in the town.
Then the linear order could be established according to locations.}
Specifically, let $<$ be a linear order over $O$,
where we henceforth interpret ``$o< o'$'' as ``the object $o$ is smaller than $o'$'' throughout.

Given a preference profile, Crawler determines the final allocation through an algorithm as follows.
Checking from the smallest object to the largest along the underlying linear order $<$, 
we identify the first agent whose favorite object is weakly smaller than her own, and let this agent leave with her favorite object.
Then, as an update, we let agents who currently hold objects located between this agent's favorite object and her own object ``crawl'' to the adjacently larger objects, and fix all other agents to their current objects.
This procedure is repeated until all agents leave.
By definition, exactly one agent leaves at each step.
Hence, the algorithm terminates in $n$ steps.

Crawler restores the compatibility of efficiency, individual rationality, and obvious strategy-proofness, given that all admissible preferences are single-peaked with respect to the underlying linear order $<$.\footnote{It is important to note that Crawler by definition does not require that the admissible preferences are single-peaked. In Example \ref{exm:indispensibility}, Crawler operates well on a domain of non-single-peaked preferences.}
Formally, a preference $P_i$ is said \textbf{single-peaked} w.r.t.~$<$ if for all $o, o' \in O$,
we have $\big[o'< o < r_1(P_i)\; \textit{or}\; r_1(P_i) < o < o'\big] \Rightarrow [o\mathrel{P_i} o']$.
Let $\mathcal{D}_<$ denote the domain that contains \emph{all} single-peaked preferences w.r.t.~$<$.
A domain $\mathcal{D}$ is called \textbf{a single-peaked domain} if 
there exists a linear order $<$ over $O$ such that $\mathcal{D} \subseteq \mathcal{D}_<$.\bigskip

\begin{proposition}[\citealt{B2019}]\label{prop:crawler}
    On the single-peaked domain $\mathcal{D}_<$, Crawler is efficient, individually rational and obviously strategy-proof.
\end{proposition}

{\centering\section{Merit-Based Fairness}\label{sec:hierarchy}}

Given a preference profile $P \in \mathcal{D}^n$, we repeatedly remove agents who prefer their endowments the most among the remaining ones.
Then, let $\bar{N}(P)$ be the set of agents who survive the removal procedure, and $\bar{O}(P)$ be the set of their endowments.
Clearly, agents excluded from $\bar{N}(P)$ get their own endowments in an individually rational allocation, and hence are inactive in the exchange. 
We call agents in $\bar{N}(P)$ \textbf{active agents} and objects in $\bar{O}(P)$ \textbf{active objects}.

We collect all active agents whose endowments are most preferred by some other active agent.
Formally, let $A(P)\equiv \left\{ i\in \bar{N}(P): ~\textrm{there exists}~j\in\bar{N}(P)~\textrm{such that}~ o_i=\max\nolimits^{P_j}\bar{O}(P) \right\}$.
We call an agent in $A(P)$ an \textbf{acclaimed agent}. 
In practical terms, an acclaimed agent is, for instance, a homeowner whose house receives the most bids, or a kidney donor whose organ is the best match for multiple patients. The merit-based principle suggests that such an agent deserves a favorable outcome in the exchange.
Note that either $\bar{N}(P) = A(P)= \emptyset$, or $|\bar{N}(P)| \geq |A(P)| \geq 2$.
%
%
For each acclaimed agent $i\in A(P)$, let $\bar{N}_i(P)\equiv
\{ j\in\bar{N}(P): \max\nolimits^{P_j}\bar{O}(P) =o_i\}$ be the set of active agents who acclaim the object $o_i$ (\textit{i.e.}, prefer $o_i$ the most within $\bar{O}(P)$).
It is evident that $i \notin \bar{N}_i(P)$ for all $i \in A(P)$.

Once an active agent $j \in \bar{N}(P)$ acclaims an object $o_i \in \bar{O}(P)$,
it indicates that $j$ is willing to exchange her own object for $o_i$, and seeks the approval of the exchange from agent $i$.
Consequently, $o_j$ is reserved for $i$ to pick, and hence $i$ would not receive any object worse than $o_j$ in the final allocation.
The axiom below formalizes this fairness concern.\medskip

\begin{definition}
    \label{def:EKA}
    Given a preference profile $P \in \mathcal{D}^n$ with $A(P) \neq \emptyset$,
    an allocation $m\in\mathcal{M}$ satisfies \textbf{equity under acclaims (EA)} at $P$ 
    if the following conditions are satisfied:
    \begin{itemize}
    \item $m(i) \in \bar{O}(P)$,\footnote{The requirement $m(i) \in \bar{O}(P)$ is automatically satisfied when the allocation $m$ is individually rational.} and 
    \item $m(i)\mathrel{R_i}o_j$~ for all $i\in A(P)$ and $j\in\bar{N}_i(P)$.
    \end{itemize}
    Correspondingly, a rule $f: \mathcal{D}^n \rightarrow \mathcal{M}$ satisfies EA
    if at each $P \in \mathcal{D}^n$ with $A(P) \neq \emptyset$, the allocation $f(P)$ satisfies EA.
\end{definition}

One can easily verify that TTC satisfies EA.

In particular, consider a preference profile $P \in \mathcal{D}^n$ such that $|A(P)| = 2$. Let $A(P) \equiv \{i,j\}$.
Observe that the set of active agents $\bar{N}(P)$ is effectively partitioned into two groups, $\bar{N}_i(P)$ and $\bar{N}_j(P)$, where $j\in \bar{N}_i(P)$ and $i\in \bar{N}_j(P)$.
As agent $i$ prefers $o_j$ to any other active object, and $o_i$ is acclaimed by agent $j$,
$i$ must receive her favorite active object $o_j$ in an allocation satisfying EA.
Symmetrically, agent $j$ gets her favorite active object $o_i = \max\nolimits^{P_j} \bar{O}(P)$.
The following definition formalizes this observation.

\begin{definition}
    \label{def:EBA}
    Given a preference profile $P \in \mathcal{D}^n$ with $|A(P)| =2$,
    an allocation $m\in\mathcal{M}$ satisfies \textbf{equity under bipartite acclaims (EBA)} at $P$ if the following condition is satisfied:
    \begin{itemize}
        \item $m(i) = \max\nolimits^{P_i} \bar{O}(P)$~ for \underline{each} $i \in A(P)$.
    \end{itemize} 
    Correspondingly, a rule $f: \mathcal{D}^n \rightarrow \mathcal{M}$ satisfies EBA
    if at each $P \in \mathcal{D}^n$ with $|A(P)| = 2$, the allocation $f(P)$ satisfies EBA.
\end{definition}

One can naturally weaken EBA by requiring at least one of the two acclaimed agents to receive her favorite active object.\footnote{For instance, given $N = \{1,2,3\}$ and the linear order $o_1<o_2<o_3$, consider a profile of single-peaked preferences $P \equiv (P_1, P_2, P_3)$ where $1$ prefers $o_3$ the most, and both $2$ and $3$ prefer $o_1$ the most.
Clearly, $1$ and $3$ are the acclaimed agents.
However, in the Crawler allocation, only $1$, not $3$, gets her favorite active object.}
This is particularly relevant when the object of one acclaimed agent is unanimously acclaimed by all other active agents.
Specifically, given a preference profile $P \in \mathcal{D}^n$ with $|A(P)|=2$,
we call an agent $i \in A(P)$ \textbf{the unanimously acclaimed agent} 
if $\bar{N}_i(P)= \bar{N}(P)\backslash \{i\}$.\footnote{Given $|A(P)|=2$ and the existence of the unanimously acclaimed agent, 
if $|\bar{N}(P)| \geq 3$, the unanimously acclaimed agent is unique; if $|\bar{N}(P)|=2$, both active agents are the unanimously acclaimed agents.}
A merit-based fair allocation in this extreme situation should be in favor of the unanimously acclaimed agent.
The next definition adopts the aforementioned weakening of EBA, and moreover insists that the unanimously acclaimed agent (if exists) should receive her favorite active object.

\begin{definition}
    Given a preference profile $P \in \mathcal{D}^n$ with $|A(P)| =2$,
    an allocation $m\in\mathcal{M}$ satisfies \textbf{weak equity under bipartite acclaims (WEBA)} at $P$ if the following two conditions are satisfied:
    \begin{itemize}
    \item $m(i)=\max\nolimits^{P_i}\bar{O}(P)$ for \underline{some} $i \in A(P)$, and 
    \item $m(i) =\max\nolimits^{P_i}\bar{O}(P)$ if $i \in A(P)$ is the unanimously acclaimed agent.
    \end{itemize}
    Correspondingly, a rule $f: \mathcal{D}^n \rightarrow \mathcal{M}$ satisfies WEBA
    if at each $P\in\mathcal{D}^n$ with $|A(P)| =2$, the allocation $f(P)$ satisfies WEBA.
\end{definition}

Finally, an allocation satisfies equity under unanimous acclaims if, whenever the unanimously acclaimed agent is present, she receives her favorite active object.

\begin{definition}
Given a preference profile $P \in \mathcal{D}^n$ with $|A(P)| =2$, 
an allocation $m\in\mathcal{M}$ satisfies \textbf{equity under unanimous acclaims (EUA)} at $P$ if the following condition is satisfied:
\begin{itemize}
\item $m(i)=\max\nolimits^{P_i}\bar{O}(P)$ if $i \in A(P)$ is the unanimously acclaimed agent.
\end{itemize}
Correspondingly, a rule $f: \mathcal{D}^n \rightarrow \mathcal{M}$ satisfies EUA
if at each $P \in \mathcal{D}^n$ with $|A(P)| =2$, the allocation $f(P)$ satisfies EUA.
\end{definition}

\citet{M2004} proposed four principles of fairness: \textit{Ex post Equality}, \textit{Reward}, \textit{Exogenous Rights} and \textit{Fitness}. Due to indivisibility, ex post equality cannot be applied to strategic exchange.
Nevertheless, EUA embodies the spirit of the remaining three principles,
since it can be interpreted as \textit{rewarding} the unanimously acclaimed agent with her favorite active object---the object she is best \emph{fitted} to utilize---based on the merit of her \emph{exogenous endowment}.

We conclude this section by highlighting the hierarchy of the four fairness axioms:
\begin{equation*}
    \text{EA}\Rightarrow \text{EBA} \Rightarrow \text{WEBA} \Rightarrow \text{EUA}.
\end{equation*}

{\centering\section{The First Cost}\label{sec:unanimous}}

We now establish the first cost: obviously strategy-proof rules that satisfy even our weakest fairness axiom cannot operate under unrestricted preferences; they can only operate under single-peaked preferences.

TTC, while strategy-proof, clearly satisfies EUA but fails to be obviously strategy-proof,
whereas Crawler, under single-peaked preferences, restores the compatibility of obvious strategy-proofness and EUA.
This suggests a cost in the design of obviously strategy-proof rules: to preserve even the least fairness requirement, exchange rules need to be bounded in a restricted environment, which clearly shrinks their domain of applicability.
Our main result, Theorem~\ref{thm:merit}, establishes that this cost is inevitable: under a mild richness condition on the preference domain, single-peakedness is not only sufficient but necessary for the existence of an efficient, individually rational and obviously strategy-proof rule that satisfies EUA.

We introduce the richness condition before the theorem.
Fixing a domain $\mathcal{D}$,
two objects $o$ and $o'$ are said \textit{connected}, denoted $o \sim o'$,
if there exist $P_i, P_i' \in \mathcal{D}$ such that
$r_1(P_i) = r_2(P_i') = o$ and $r_1(P_i') = r_2(P_i) = o'$.
Domain $\mathcal{D}$ is called a \textbf{path-connected} domain
if for any two distinct objects $o, o' \in O$,
there exists a sequence of non-repeated objects $(x^1, \dots, x^q)$ such that
$x^1 = o$, $x^q = o'$ and $x^k \sim x^{k+1}$ for all $k = 1, \dots, q-1$.
Furthermore, to impose sufficient diversity on preferences,
we require the domain $\mathcal{D}$ to contain at least one pair of two completely reversed preferences, \textit{i.e.}, there exist $\underline{P}_i, \overline{P}_i \in \mathcal{D}$ such that $\big[o\mathrel{\underline{P}_i}o'\big] \Leftrightarrow \big[o' \mathrel{\overline{P}_i} o\big]$.
Henceforth, a path-connected domain that contains a pair of completely reversed preferences is simply called a \textbf{rich domain}.\footnote{We make clear the following three points on the richness condition.
    First, since connectedness imposes no requirement on the rankings of objects beyond the top-two positions, 
    path-connectedness covers various well-studied preference domains, like the unrestricted domain, the single-peaked domain, the domain of separable preferences \citep{BSZ1991,LS1999}, the domain of multidimensional single-peaked preferences \citep{BGS1993}, etc.
    Second, a rich domain can be sparse. For instance, given $n$ objects, the single-peaked domain has $2^{n-1}$ preferences, while the cardinality of a rich single-peaked domain can be as small as $2(n-1)$.
    Third, \citet{T2023GEB} showed that TTC satisfies obvious strategy-proofness on the domain of single-dipped preferences, which however is excluded by our richness condition.
    In particular, path-connectedness allows each object to be top-ranked in some preference of the domain, whereas 
    in each single-dipped preference, the top-ranked object must be either one of the two extreme objects on the underlying linear order.
    The imposition of single-dippedness indeed does not enlarge the scope of designing exchange rules, as \citet{T2023GEB} proved that TTC is uniquely characterized by efficiency, individual rationality and strategy-proofness,
    whereas on the single-peaked domain, various exchange rules beyond TTC and Crawler preserve to be strategy-proof or obviously strategy-proof.
    }

\bigskip

\begin{theorem}\label{thm:merit}
    Let $\mathcal{D}$ be a rich domain.
    There exists a rule $f:\mathcal{D}^n\rightarrow\mathcal{M}$ satisfying efficiency, individual rationality, obvious strategy-proofness, and EUA if and only if $\mathcal{D}$ is a single-peaked domain.
\end{theorem}

The proof of Theorem \ref{thm:merit} is in Appendix \ref{app:merit}.
Here, we give a brief intuition of it.

By Proposition~\ref{prop:crawler}, we have an obviously strategy-proof exchange rule--Crawler defined on the single-peaked domain. Rather than verifying directly that Crawler satisfies EUA, we show more generally that on the single-peaked domain, \emph{all} efficient, individually rational and strategy-proof rules, including Crawler, endogenously satisfy EUA.

For the necessity part of Theorem~\ref{thm:merit}, we illustrate the proof strategy with a simple example of three objects.
Let $N = \{1, 2, 3\}$ and $O = \{o_1, o_2, o_3\}$.
All six preferences of the unrestricted domain $\mathcal{P}$ are specified in Table \ref{tab:unrestricted}.
\begin{table}[h]
\centering
        \begin{tabular}{cccccc}
            \hline\hline\rule[0mm]{0mm}{4mm}
            $P_i^1$ & $P_i^2$ & $P_i^3$ & $P_i^4$ & $P_i^5$ & $P_i^6$ \\ \hline
            $o_1$ & $o_1$ & $o_2$  & $o_3$ & $o_2$ & $o_3$\\[-0.2em]
            $o_3$ & $o_2$ & $o_1$  & $o_2$ & $o_3$ & $o_1$\\[-0.2em]
            $o_2$ & $o_3$ & $o_3$  & $o_1$ & $o_1$ & $o_2$ \\[0.2em] \hline\hline
        \end{tabular}
\caption{The unrestricted domain $\mathcal{P}$}\label{tab:unrestricted}
\end{table}
Let $\mathcal{D} \subseteq \mathcal{P}$ be a rich domain and let $f: \mathcal{D}^3 \rightarrow \mathcal{M}$ be a rule satisfying the properties mentioned in Theorem~\ref{thm:merit}.
By path-connectedness of $\mathcal{D}$, we assume w.l.o.g.~that $o_1 \sim o_2$ and $o_2 \sim o_3$,
which implies that preferences $P^2, P^3, P^4,P^5$ are included in $\mathcal{D}$.
Note that these four preferences are single-peaked w.r.t.~the underlying linear order $o_1 < o_2 < o_3$.
Therefore, in this example, it suffices to show that $P^1$ and $P^6$ are excluded from the domain $\mathcal{D}$.
Suppose by contradiction that $P^1 \in \mathcal{D}$ (an analogous argument works for $P^6$).
Consequently, when restricting attention to sub-domains
$\mathcal{D}_1 = \{P_1^3, P_1^4\}$,
$\mathcal{D}_2 = \{P_2^1, P_2^4\}$ and
$\mathcal{D}_3= \{P_3^2, P_3^3\}$,
we show that efficiency, individual rationality, strategy-proofness, and EUA together force the rule $f$ to deliver TTC allocations at all related preference profiles.
However, analogous to the proof of Proposition~5 in \citet{L2017}, this leads $f$ to a violation of obvious strategy-proofness.
This simple example reveals ``local single-peakedness'' \citep[equivalently, the never-bottom value restriction of][]{S1966} over three connected objects, a condition that eventually expands to ``global single-peakedness'' via transitivity along the sequences provided by path-connectedness.

We conclude this section with two remarks.
First, the example below shows the indispensability of EUA in pinning down single-peakedness.
Specifically, we construct a rich but non-single-peaked domain on which Crawler remains to be efficient, individually rational and obviously strategy-proof, yet violates EUA.

\begin{example}\label{exm:indispensibility}\rm
    Let $N = \{1, 2, 3\}$ and $O = \{o_1, o_2, o_3\}$, where $o_1 < o_2 < o_3$.
    Recall the unrestricted domain $\mathcal{P}$ specified in Table \ref{tab:unrestricted}.
    According to the underlying linear order $<$, the single-peaked domain is $\mathcal{D}_<=\{ P_i^2,P_i^3,P_i^4,P_i^5 \}$.
    Let $\mathcal{D} = \{P_i^1, P_i^2,P_i^3,P_i^4,P_i^5\}$ be a domain of five preferences.
    Since $\mathcal{D}_<$ is strictly contained in $\mathcal{D}$,
    it is clear that $\mathcal{D}$ is a rich but non-single-peaked domain.
    
    It can be shown that Crawler $\mathscr{C}: \mathcal{D}^3 \rightarrow \mathcal{M}$ according to $<$ is an efficient and individually rational rule that can be OSP-implemented by a millipede game and the greedy-strategy plans of \cite{PT2023} (see the detailed verification in Appendix \ref{app:indispensibility}).
    However, Crawler $\mathscr{C}$ violates EUA here.
    For instance, consider the preference profile specified below:
    
     \begin{table}[h]
        \centering
        \begin{tabular}{ccc}
            \hline\hline \rule[0mm]{0mm}{4mm}
            $P_1^4$ & $P_2^2$ & $P_3^1$ \\
            \hline
            $o_3$  & $o_1$  & $o_1$ \\
            $o_2$  & $o_2$  & $o_3$ \\
            $o_1$  & $o_3$  & $o_2$ \\[0.2em]
            \hline\hline
        \end{tabular}
    \caption{Preference profile $P \equiv (P_1^4, P_2^2, P_3^1)$}\label{tab:preferenceprofile}
    \end{table}
    
    \noindent
    It is clear that $\bar{N}(P) = \{1, 2, 3\}$, $\bar{O}(P) = \{o_1, o_2, o_3\}$, $A(P) = \{1,3\}$, and
    agent $1$ is the unanimously acclaimed agent at $P$.
    However, agent 1 does not receive her favorite active object at the Crawler allocation, \textit{i.e.}, $\mathscr{C}_1(P) = o_2 \neq o_3 =\max\nolimits^{P_1^4} \bar{O}(P)$.
    \hfill$\Box$
\end{example}

The second remark situates Theorem~\ref{thm:merit} within the broader literature on mechanism design with single-peaked preferences.
Both strategy-proof and obviously strategy-proof mechanisms under single-peaked preferences have been widely explored \citep[\emph{e.g.},][]{M1980,S1991,BBM2020,AMN2020,AMN2023,B2019}.
In the past decade, a growing line of work has investigated how essential single-peaked preferences are for allowing fairness in designing strategy-proof mechanisms in the models of strategic voting \citep{CSS2013} and division \citep{CMS2025}.
This can be traced back to ``Gul's Conjecture'' \citep[see Section 6.5.2 of][]{B2011} in the late 1980s, in which Faruk Gul speculated that in a strategic voting model, according to a strategy-proof social choice function,
the alternatives in the range are able to be embedded in a grid such that all preferences turn out to be single-peaked corresponding to the embedding.
Our Theorem~\ref{thm:merit} contributes to this literature by showing that in the strategic exchange model, single-peakedness arises as a consequence of the existence of an obviously strategy-proof rule that satisfies a merit-based fairness criterion.

{\centering\section{The Second Cost}\label{sec:upperbound}}

We next identify the second cost: even under single-peaked preferences, the level of attainable fairness is substantially reduced in the design of obviously strategy-proof rules.
Specifically, we show that EBA is not satisfied by any efficient, individually rational and obviously strategy-proof rule, whereas WEBA is attainable by a new rule.

We begin with an example which shows that Crawler violates both EBA and WEBA.

\begin{example}\rm
    \label{eg:merit-crawler}
    Let $N=\{1,2,3,4\}$ and $O=\{o_1,o_2,o_3,o_4\}$, where $o_1< o_2< o_3< o_4$.
    Consider a profile of single-peaked preferences $P \equiv (P_1, P_2, P_3, P_4) \in \mathcal{D}_<^4$ specified in Table \ref{tab:4411}.
    \begin{table}[h]
    \centering
            \begin{tabular}{cccc}
                \hline\hline\rule[0mm]{0mm}{4mm}
                $P_1$ & $P_2$ & $P_3$ & $P_4$ \\ \hline
                \rule[0mm]{0mm}{4mm}$o_4$ & $\boxed{o_4}$ & $\boxed{o_1}$ & $o_1$ \\[-0.2em]
                $\boxed{o_3}$ & $o_3$ & $o_2$ & $\boxed{o_2}$ \\[-0.2em]
                $o_2$ & $o_2$ & $o_3$ & $o_3$ \\[-0.2em]
                $o_1$ & $o_1$ & $o_4$ & $o_4$\\[0.2em] \hline\hline
            \end{tabular}
            \caption{A profile of single-peaked preferences}\label{tab:4411}
    \end{table}
    
    All four agents are active at $P$, $A(P) = \{1,4\}$, and there is no unanimously acclaimed agent.
    Although both 1 and 4 are the acclaimed agents at $P$, none of them at the Crawler allocation $\mathscr{C}(P)$, illustrated by the boxes in Table \ref{tab:4411}, gets her favorite active object,
    which indicates a violation of the requirements of both EBA and WEBA.
    \hfill$\Box$
\end{example}

The theorem below further shows that on the single-peaked domain, no efficient, individually rational and obviously strategy-proof rule satisfies EBA.

\begin{theorem}\label{thm:impossibility}
    Let $n\geq4$.
    On the single-peaked domain $\mathcal{D}_<$~, no efficient, individually rational and obviously strategy-proof rule satisfies EBA.\footnote{When $n = 3$, \citet{B2019} showed that TTC, which clearly satisfies EBA, is obviously strategy-proof on the domain of single-peaked preferences.}
\end{theorem}

We nevertheless verify that it is possible to improve Crawler's fairness performance of  via a new rule, called Designator, 
which blends the sequential structure of Crawler with the merit-based fairness embodied by TTC, to preserve obvious strategy-proofness and meanwhile guarantee WEBA.

To define this rule, we first introduce some new notation for ease of presentation.
A sub-allocation $\bar{m}$ is a one-to-one mapping from a subset of agents $N'\subseteq N$ to a subset of objects $O'\subseteq O$ such that $|N'|=|O'|$, where $\bar{m}(i)$ denotes the object in $O'$ assigned to agent $i\in N'$.
Let $N_{\bar{m}}$ and $O_{\bar{m}}$ denote respectively the set of agents and the set of objects involved in the sub-allocation $\bar{m}$.
Hence, an allocation $m \in \mathcal{M}$ is a special sub-allocation where $N_m=N$.
We sometimes write a sub-allocation as a set of agent-object pairs.
For instance, the endowment $e$ can be written as $e=\{ (1,o_1),\dots,(n,o_n) \}$.
Fixing a linear order $<$ over $O$ and two objects $o',o''\in O$ such that $o' < o''$, let $[o',o'')\equiv \{ o\in O: o'\leq o<o'' \}$ denote the interval of objects that are weakly larger than $o'$ and smaller than $o''$.
Moreover, given a sub-allocation $\bar{m}$ and two objects $o', o'' \in O_{\bar{m}}$,
we say that $o''$ is adjacently larger than $o'$ in $O_{\bar{m}}$ if $o'<o''$, and there exists no $o \in O_{\bar{m}}$ such that $o' < o < o''$.

\begin{definition}\label{def:designator}
    \textbf{Designator} $\mathscr{D}:\mathcal{D}_<^n\rightarrow \mathcal{M}$ is a rule
    such that at each $P\in \mathcal{D}_<^n$,
    the allocation $\mathscr{D}(P)$ is determined through the algorithm below.
    \begin{description}
        \item[Step $0$ (Preparation):]
        Let $\mathscr{D}_{i}(P) = o_i$ for all $i \in N\backslash\bar{N}(P)$ and
        $\bar{m}^0\equiv \big\{(i, o_i): i \in \bar{N}(P)\big\}$.
        The algorithm terminates if $\bar{N}(P) = \emptyset$.
        Otherwise, proceed to the next step.
        
        \item[Step $\bm{s} \geq 1$:]
        ~~
        
        \begin{description}
            \item[Identification] Checking from the smallest object to the largest in $N_{\bar{m}^{s-1}}$ according to the linear order $<$, identify the first agent, denoted $i^s$, whose most preferred object in $O_{\bar{m}^{s-1}}$ is weakly smaller than her currently held object, i.e., $\max\nolimits^{P_{i^s}} O_{\bar{m}^{s-1}} \leq \bar{m}^{s-1}(i^s)$. Denote $\underline{o}^s \equiv \max\nolimits^{P_{i^s}} O_{\bar{m}^{s-1}}$ and
                $o^s \equiv \bar{m}^{s-1}(i^s)$.
            
            \item[Allocation] Let $\mathscr{D}_{i^s}(P) = \underline{o}^s$, i.e., let $i^s$ take $\underline{o}^s$ and leave.
            
            \item[Recognition] Identify the acclaimed agent $j^s \in A(P)$ such that $i^s \in \bar{N}_{j^s}(P)$, and check whether $j^s$ remains in $N_{\bar{m}^{s-1}}$ and holds an object in the interval $[\underline{o}^s, o^s)$.
                If $j^s\in N_{\bar{m}^{s-1}}$ and $\bar{m}^{s-1}(j^s) \in [\underline{o}^s, o^s)$, call $j^s$ ``the designated agent'', and denote $\bar{o}^s \equiv \bar{m}^{s-1}(j^s)$.
                
            \item[Updating] If $|\bar{m}^{s-1}|=1$, the algorithm terminates.
            Otherwise, update $\bar{m}^{s-1}$ to $\bar{m}^s$ as follows:

            \begin{description}
                \item[\sc Trivial updating:] If $\underline{o}^s=o^s$, let $\bar{m}^s=\bar{m}^{s-1}\backslash\{ (i^s, o^s) \}$.
                \item[\sc Designating updating:] If $\underline{o}^s< o^s$, $j^s\in N_{\bar{m}^{s-1}}$ and $\bar{o}^s \in [\underline{o}^s, o^s)$,
                then $j^s$ is designated to inherit $o^s$, i.e., $\bar{m}^s(j^s)=o^s$, let every agent of $N_{\bar{m}^{s-1}}$ whose current object is in the interval $[\underline{o}^s,\bar{o}^s)$, if nonempty, ``crawl'' to the adjacently larger object in $O_{\bar{m}^{s-1}}$, and fix all other agents of $N_{\bar{m}^{s-1}}$ to their current objects.

                \item[\sc Crawling updating:] Otherwise, let every agent of $N_{\bar{m}^{s-1}}$ whose current object is in the interval $[\underline{o}^s, o^s)$ ``crawl'' to the adjacently larger object in $O_{\bar{m}^{s-1}}$, and fix all other agents of $N_{\bar{m}^{s-1}}$ to their current objects.
                
            \end{description}
            After the updating, the algorithm proceeds to the next step.
        \end{description}
    \end{description}    
\end{definition}

Of course, at each preference profile $P \in \mathcal{D}_<^n$, the algorithm terminates at Step $|\bar{N}(P)|$.
It is worth mentioning that if both parts of Recognition and Designating updating in each Step $s\geq 1$ are removed,
the algorithm delivers the Crawler allocation at each preference profile.

The following illustrates Designator applied to the preference profile of Example~\ref{eg:merit-crawler}.

\begin{example}\label{exm:designator}\rm
    Recall the preference profile $P$ specified in Table \ref{tab:4411}, where $A(P) = \{1,4\}$.
    The Designator allocation $\mathscr{D}(P)$ is determined through the procedure below.\medskip
    
\noindent
\textbf{Step} 0 (\textbf{Preparation}): We have $\bar{N}(P) = N$ and $\bar{m}^0 = e$. The algorithm proceeds to Step 1.\medskip
        
\noindent
\textbf{Step} 1: 
(i) Checking from $o_1$ to $o_4$, identify $i^1 = 3$, the first agent in $N_{\bar{m}^0}$ whose favorite object in $O_{\bar{m}^0}$ is weakly smaller than her currently held object.  
Denote $\underline{o}^1 \equiv \max\nolimits^{P_3} O_{\bar{m}^0} = o_1$ and $o^1 \equiv \bar{m}^0(3) = o_3$.      
(ii) Let $\mathscr{D}_3(P)=\underline{o}^1 = o_1$.
(iii) Identify the acclaimed agent $1 \in A(P)$ such that $3 \in \bar{N}_1(P)$.
Agent $1$ is recognized as the designated agent as $1 \in N_{\bar{m}^0}$ and $\bar{m}^0(1) = o_1 \in [o_1, o_3) = [\underline{o}^1, o^1)$.
(iv) Apply the designating updating: agent $1$ is designated to inherit $o^1 = o_3$, while fixing both 2 and 4 to their objects in $\bar{m}^0$. Thus, we have $\bar{m}^1 = \big\{(2, o_2), (1, o_3), (4,o_4)\big\}$, and the algorithm proceeds to Step 2.
   \medskip
   
\noindent     
\textbf{Step} 2: 
(i) Checking from $o_2$ to $o_4$, identify $i^2 = 4$. Denote $\underline{o}^2 = \max\nolimits^{P_4}O_{\bar{m}^1} = o_2$ and $o^2 \equiv \bar{m}^1(4) =o_4$.
(ii) Let $\mathscr{D}_4(P)=\underline{o}^2 = o_2$.
(iii) Identify the acclaimed agent $1 \in A(P)$ such that $4 \in \bar{N}_1(P)$.
Agent $1$ is recognized as the designated agent as $1 \in N_{\bar{m}^1}$ and $\bar{m}^1(1) = o_3 \in [o_2, o_4) = [\underline{o}^2, o^2)$.      
(iv) Apply the designating updating: agent $1$ is designated to inherit $o^2 = o_4$, and agent 2 crawls to the adjacently larger object $o_3$. Thus, we have $\bar{m}^2 = \big\{(2, o_3), (1,o_4)\big\}$, and the algorithm proceeds to Step 3.\medskip
       
\noindent 
\textbf{Step 3}: 
(i) Checking from $o_3$ to $o_4$, identify $i^3 =1$. Denote $\underline{o}^3 \equiv \max\nolimits^{P_1}O_{\bar{m}^2} = o_4$ and $o^3 \equiv \bar{m}^3(1) = o_4$.
(ii) Let $\mathscr{D}_1(P)=\underline{o}^3 = o_4$.
(iii) Identify the acclaimed agent $4 \in A(P)$ such that $1 \in \bar{N}_4(P)$.
Agent $4$ is not recognized as the designated agent as $4 \notin N_{\bar{m}^2}$.
(iv) By applying the trivial updating, we have $\bar{m}^3 = \{(2,o_3)\}$.
Then, the algorithm proceeds to Step 4.\medskip
        
\noindent
\textbf{Step} 4:  Agent $2$ gets $o_3$ and the algorithm terminates.
\medskip

According to the procedure above, we have the designator allocation $\mathscr{D}(P)$ specified in the boxes of the following table.
\begin{table}[h]
    \centering
            \begin{tabular}{cccc}
                \hline\hline\rule[0mm]{0mm}{4mm}
                $P_1$ & $P_2$ & $P_3$ & $P_4$ \\ \hline
                \rule[0mm]{0mm}{4mm}$\boxed{o_4}$ & $o_4$ & $\boxed{o_1}$ & $o_1$ \\[-0.2em]
                $o_3$ & $\boxed{o_3}$ & $o_2$ & $\boxed{o_2}$ \\[-0.2em]
                $o_2$ & $o_2$ & $o_3$ & $o_3$ \\[-0.2em]
                $o_1$ & $o_1$ & $o_4$ & $o_4$\\[0.2em] \hline\hline
            \end{tabular}
            \caption{The designator allocation $\mathscr{D}(P)$}
    \end{table}

\noindent
One can easily see that $\mathscr{D}(P)$ satisfies WEBA, as the acclaimed agent 1 gets her favorite active object. \hfill$\Box$  
\end{example}

The following theorem formalizes these properties.

\begin{theorem}
    \label{thm:designator}
    Designator $\mathscr{D}: \mathcal{D}_<^n \rightarrow \mathcal{M}$ satisfies efficiency, individual rationality, obvious strategy-proofness, and WEBA.
\end{theorem}

The proof of Theorem~\ref{thm:designator} is provided in Appendix~\ref{app:designator}.
We highlight two key features of Designator that explain its compliance with WEBA and obvious strategy-proofness.

First, at each preference profile with exactly two acclaimed agents, Designator always favors the one with the smaller endowment, \textit{that is}, at each $P \in \mathcal{D}_<^n$ with $A(P) \equiv \{i,j\}$, $[o_i < o_j]\Rightarrow \big[\mathscr{D}_i(P) = o_j = \max\nolimits^{P_i} \bar{O}(P)\big]$ (see for instance Example \ref{exm:designator} and the formal verification in Lemma \ref{lem:priority} of Appendix~\ref{app:designator}).\footnote{Even when $j$ turns out to be the unanimously acclaimed agent, 
this key feature of $\mathcal{D}(P)$ is consistent with EUA,
because in this case, agent $j$ receives $o_i$, her favorite active object, at Step 1 of the algorithm.}

Second, the designated agent at a given step may inherit a strictly worse object. This may appear at odds with the requirement of obvious strategy-proofness. However, the algorithm of Designator ensures that this agent eventually receives an object no worse than any object she temporarily held during the procedure.
We call this property ``dynamic individual rationality'' (see the detailed explanation in Appendix \ref{app:DynamicIR}), which is critical for ensuring Designator to be obviously strategy-proof.

We summarize this section by introducing a table below that contrasts the allocations produced by TTC, Crawler, and Designator at the preference profile of Example~\ref{eg:merit-crawler}.
TTC satisfies EBA but not obvious strategy-proofness; Crawler satisfies obvious strategy-proofness and EUA, but violates WEBA; Designator satisfies both obvious strategy-proofness and WEBA.

\begin{table}[ht]
    \centering
    \begin{tabular}{cc|ccc}
        \hline\hline
        \multicolumn{2}{c|}{Agent}  & TTC & Crawler & Designator \\
        \hline
        \multicolumn{2}{c|}{1} & $o_4$ & $o_3$ & $o_4$ \\
        \multicolumn{2}{c|}{2}  & $o_3$ & $o_4$ & $o_3$ \\
        \multicolumn{2}{c|}{3}  & $o_2$ & $o_1$ & $o_1$ \\
        \multicolumn{2}{c|}{4}  & $o_1$ & $o_2$ & $o_2$ \\[0.2em]
        \hline
        \multicolumn{2}{c|}{Fairness}& EBA  & EUA & WEBA \\[0.1em]
        \hline
        \multicolumn{2}{c|}{Incentive Property}& Strategy-proof  & Obviously Strategy-proof & Obviously Strategy-proof \\[0.1em]
        \hline\hline
    \end{tabular}
    \caption{A comparison of rules at the preference profile in Example~\ref{eg:merit-crawler}.}
    \label{tab:comparison}
\end{table}

{\centering\section{Conclusion}\label{sec:conclusion}}

In this paper, we propose a hierarchy of merit-based fairness notions:
$\textrm{EA} \Rightarrow \textrm{EBA} \Rightarrow \textrm{WEBA} \Rightarrow \textrm{EUA}$,
which require an exchange rule to fairly reward the acclaimed agents for their provision of desired objects to the economy.
Together, our three theorems which investigate the compatibility of obvious strategy-proofness and our fairness requirements, identify two clear costs of strengthening strategy-proofness to obvious strategy-proofness in the design of exchange rules. 
Specifically, the first cost is a narrowing of the applicability of obviously strategy-proof exchange rules, namely, TTC is strategy-proof on the unrestricted domain while obviously strategy-proof rules satisfying the weakest axiom in our hierarchy can be applied only when the preference restriction of single-peakedness can be plausibly assumed.
The second cost is a substantial reduction in fairness, namely, 
WEBA is a fairness upper bound in our hierarchy that is attainable by obviously strategy-proof rules defined on the single-peaked domain, which is substantially weaker than the fairness standard delivered by TTC.

This paper makes four specific contributions to the literature on strategic exchange.
First, we propose a novel hierarchy of merit-based fairness axioms, providing the first formal analysis of fairness in the classical exchange rules.
Second, we systematically identify two unambiguous costs of strengthening strategy-proofness to obvious strategy-proofness.
Third, we theoretically justify the role of single-peaked preferences for combining obvious strategy-proofness with fairness in the exchange setting, linking our result to the broader literature on the salience of single-peakedness in incentive compatible mechanism design.
Fourth, we introduce a new obviously strategy-proof rule: Designator that outperforms the existing obviously strategy-proof rules in fairness.

Our fairness hierarchy provides a new perspective for studying fairness in strategic exchange, and can be applied to further investigations.
First, the hierarchy can be refined. One natural direction is to formulate fairness criteria that weigh the intensity of acclamation: for instance, an agent whose object is the second-best for everyone else may deserve greater priority than one whose object is the favorite of a single agent. 
Second, extending the merit-based fairness framework to settings with monetary transfers, such as strategic exchange with transfers \citep{M2001}, two-sided matching with transfers \citep{SS1971,DG1985}, and voting with money \citep{R1979,MMP2018}, is a promising avenue for connecting the fairness implications identified here to a broader class of mechanism design problems. We leave these investigations for future work.

\appendix

{\centering\section*{Appendix}}

{\centering\section{Proof of Theorem~\ref{thm:merit}} \label{app:merit}}

\noindent
\textbf{Sufficiency}.
Let $\mathcal{D}$ be a single-peaked domain, \textit{i.e.}, $\mathcal{D} \subseteq \mathcal{D}_<$ for some linear order $<$ over $O$.
For the sufficiency part, it suffices to show that on the single-peaked domain $\mathcal{D}_<$,
there exists an efficient, individually rational and obviously strategy-proof rule that satisfies EUA.
Rather than verifying directly that Crawler satisfies EUA, we show that on the single-peaked domain $\mathcal{D}_<$, all efficient, individually rational and strategy-proof rules endogenously satisfy EUA.

Let $f: \mathcal{D}_<^n\rightarrow \mathcal{M}$ be an efficient, individually rational and strategy-proof rule.
Given a profile $P \in \mathcal{D}_<^n$ with $|A(P)| = 2$, let $A(P) \equiv \{i,j\}$ and $i$ be the unanimously acclaimed agent.
Clearly, $o_i, o_j \in \bar{O}(P)$, $\bar{N}_i(P) = \bar{N}(P)\backslash \{i\}$ and $\max\nolimits^{P_i}\bar{O}(P) \equiv o_j$.
To prove that the allocation $f(P)$ satisfies EUA, we show $f_i(P) = o_j$.
Suppose by contradiction that $f_i(P) \neq o_j$.
We assume w.l.o.g.~that $o_j < o_i$.
By definition, there exists a single-peaked preference $P_i' \in \mathcal{D}_<$ such that
$r_1(P_i') = o_j$ and $[o< o_j\; \textrm{or}\; o_i< o] \Rightarrow [o_i\mathrel{P_i'}o]$.
For notational convenience, let $P' \equiv (P_i', P_{-i})$ and $f_i(P') \equiv o_{\ell}$.
Note that $\bar{N}(P) = \bar{N}(P')$, and individual rationality implies $o_j \leq o_{\ell} \leq o_i$.
If $o_{\ell} = o_j$, agent $i$ would manipulate at $P$ via $P_i'$.
Hence, it must be $o_j < o_{\ell} \leq o_i$.
Consequently, $f_k(P') = o_j$ for some $k \in N\backslash \{i\}$.
We claim $k \in \bar{N}(P)$. Suppose not, \textit{i.e.}, $k \in N\backslash\bar{N}(P) = N\backslash \bar{N}(P')$.
This, by individual rationality, immediately implies $f_k(P') = o_k$.
Hence, we have $k = j \in A(P) \subseteq \bar{N}(P)$, a contradiction.
Since $\bar{N}_i(P) = \bar{N}(P)\backslash \{i\}$, $k \in \bar{N}(P)$ implies $k \in \bar{N}_i(P)$. Hence, we know $\max\nolimits^{P_k}\bar{O}(P) = o_i$ and $o_i \mathrel{P_k} o_j$,
which, by $o_j < o_{\ell} \leq o_i$ and single-peakedness, implies $o_{\ell} \mathrel{P_k} o_j$.
Now, by letting $i$ and $k$ exchange their allocations at $f(P')$ and fixing all others' allocations at $f(P')$,
we have a new allocation that Pareto dominates $f(P')$ at $P'$.
This contradicts the efficiency of $f(P')$.
Therefore, it must be $f_i(P) = o_j$.
This proves the sufficiency part of the Theorem.\medskip

\noindent
\textbf{Necessity}.
We first introduce a new notion and a lemma that will be used in the subsequent verification.
Given three objects $x, y, z \in O$,
let $\overrightharpoon{(x,y,z)}$ denote the preference restriction
that whenever $x$ outranks both $y$ and $z$ in a preference $P_i \in \mathcal{D}$,
$y$ is ranked above $z$, \textit{i.e.},
$\big[x = \max\nolimits^{P_i}\{x, y, z\}\big] \Rightarrow \big[y \mathrel{P_i}z\big]$.
Correspondingly, let $\mathcal{R}(\mathcal{D})$ collect all such preference restrictions.
Moreover, let the ternary relation $\overline{(x, y, z)}$ denote $\overrightharpoon{(x,y,z)} \in \mathcal{R}(\mathcal{D})$ and $\overrightharpoon{(z,y,x)} \in \mathcal{R}(\mathcal{D})$.
Then, let $\mathcal{B}(\mathcal{D})$ be the set of all such ternary relations.

\begin{lemma}\label{lem:transitivity}
    All ternary relations of $\mathcal{B}(\mathcal{D})$ are transitive, \textit{i.e.},
    given four distinct objects $x, y, z, o \in O$,
    $\left[\overline{(x, y, z)}, \overline{(y, z, o)} \in \mathcal{B}(\mathcal{D})\right]
    \Rightarrow \left[\overline{(x, y, o)}, \overline{(x, z, o)} \in \mathcal{B}(\mathcal{D})\right]$.
\end{lemma}

\begin{proof}
    Given four distinct objects $x, y, z, o \in O$,
    let $\overline{(x, y, z)}, \overline{(y, z, o)} \in \mathcal{B}(\mathcal{D})$.
    Thus, we have the restrictions $ \overrightharpoon{(x,y,z)}, \overrightharpoon{(z,y,x)},
    \overrightharpoon{(y,z,o)}, \overrightharpoon{(o,z,y)} \in \mathcal{R}(\mathcal{D})$.
    Suppose $\overline{(x, y, o)} \notin \mathcal{B}(\mathcal{D})$.
    Thus, either $\overrightharpoon{(x,y,o)} \notin \mathcal{R}(\mathcal{D})$
    or $\overrightharpoon{(o,y,x)} \notin \mathcal{R}(\mathcal{D})$ holds.
    
    If $\overrightharpoon{(x,y,o)} \notin \mathcal{R}(\mathcal{D})$,
    there exists $P_i \in \mathcal{D}$ such that
    $x = \max\nolimits^{P_i}\{x, y, o\}$ and $o \mathrel{P_i} y$.
    According to the ranking of $z$ in $P_i$, there are three cases to consider:
    (i) $z\mathrel{P_i}x$,
    or (ii) $x \mathrel{P_i} z$ and $z \mathrel{P_i} y$,
    or (iii) $y \mathrel{P_i} z$.
    In case (i), $z = \max\nolimits^{P_i}\{z,y,x\}$ and $x\mathrel{P_i} y$,
    which contradicts the restriction $\overrightharpoon{(z,y,x)}$.
    In case (ii), $x = \max\nolimits^{P_i}\{x,y,z\}$ and $z\mathrel{P_i} y$,
    which contradicts the restriction $\overrightharpoon{(x,y,z)}$.
    In case (iii), $o = \max\nolimits^{P_i}\{o, z, y\}$ and $y\mathrel{P_i} z$,
    which contradicts the restriction $\overrightharpoon{(o,z,y)}$.
    Symmetrically, we can rule out $\overrightharpoon{(o,y,x)} \notin \mathcal{R}(\mathcal{D})$.
    Therefore, $\overline{(x,y,o)} \in \mathcal{B}(\mathcal{D})$.
    By a symmetric argument, we can also show $\overline{(x, z, o)} \in \mathcal{B}(\mathcal{D})$.
\end{proof}

\medskip
Henceforth, let $f: \mathcal{D}^n \rightarrow \mathcal{M}$ be a rule satisfying efficiency, individual rationality, obvious strategy-proofness, and EUA,
where $\mathcal{D}$ is a rich domain.

\begin{lemma}\label{lem:localSP}
    Fix two distinct objects $o_j, o_k \in O$ such that $o_j \sim o_k$.
    Given $o_i \in O\backslash \{o_j,o_k\}$,
    we have either $\overrightharpoon{(o_i, o_j, o_k)} \in \mathcal{R}(\mathcal{D})$ or
    $\overrightharpoon{(o_i, o_k, o_j)} \in \mathcal{R}(\mathcal{D})$.
\end{lemma}

\begin{proof}
    Suppose by contradiction that $\overrightharpoon{(o_i,o_j,o_k)} \notin \mathcal{R}(\mathcal{D})$ and $\overrightharpoon{(o_i, o_k, o_j)} \notin \mathcal{R}(\mathcal{D})$.
    Thus, there exists a preference $P_{\ell}^1 \in \mathcal{D}$ such that
    $\max\nolimits^{P_{\ell}^1}\{o_i, o_j, o_k\} = o_i$ and $o_k \mathrel{P_{\ell}^1} o_j$, and
    there exists $P_{\ell}^2 \in \mathcal{D}$ such that
    $\max\nolimits^{P_{\ell}^2}\{o_i, o_k, o_j\} = o_i$ and $o_j \mathrel{P_{\ell}^2} o_k$.
    The subscript $\ell$ here can be either agent $i$, $j$ or $k$.
    Furthermore, since $o_j \sim o_k$,
    we have two preferences $P_{\ell}^3, P_{\ell}^4 \in \mathcal{D}$ such that
    $r_1(P_{\ell}^3) = r_2(P_{\ell}^4) = o_j$ and $r_1(P_{\ell}^4) = r_2(P_{\ell}^3) = o_k$.
    According to $P_{\ell}^1, P_{\ell}^2, P_{\ell}^3, P_{\ell}^4$,
    by eliminating all objects other than $o_i, o_j$ and $o_k$, we induce a domain $\widehat{\mathcal{D}}$ of four preferences in Table \ref{tab:induceddomain}.
    \begin{table}[t]

        \centering
        \begin{tabular}{cccc}
            \hline\hline \rule[0mm]{0mm}{5mm}
            $\widehat{P}_{\ell}^1$ & $\widehat{P}_{\ell}^2$  & $\widehat{P}_{\ell}^3$ & $\widehat{P}_{\ell}^4$ \\
            \hline
            $o_i$                   & $o_i$                  & $o_j$                  & $o_k$  \\[-0.2em]
            $o_k$                   & $o_j$                  & $o_k$                  & $o_j$  \\[-0.2em]
            $o_j$                   & $o_k$                  & $o_i$                  & $o_i$\\[0.2em]
            \hline\hline
        \end{tabular}
        \caption{Induced domain $\widehat{\mathcal{D}}$}\label{tab:induceddomain}
    \end{table}
    For each agent $v \in N \backslash \{i,j,k\}$, 
    we can fix a preference $\widetilde{P}_v \in \mathcal{D}$ such that $r_1(\widetilde{P}_v) = o_v$.
    Given $P_i, P_j, P_k \in \{P_{\ell}^1, P_{\ell}^2, P_{\ell}^3, P_{\ell}^4\}$,
    by individual rationality, it is clear that $f_{\ell}\big(P_i,P_j,P_k, \widetilde{P}_{-\{i,j,k\}}\big) \in \{o_i,o_j,o_k\}$ for all $\ell \in \{i,j,k\}$.
    Then, we can construct a rule $\hat{f}$ that allocates objects $o_i, o_j,o_k$ to agents $i,j,k$ according to the induced preferences of $\widehat{\mathcal{D}}$: for each preference profile  $(\widehat{P}_i, \widehat{P}_j, \widehat{P}_k) \in \widehat{\mathcal{D}}^3$,
    after identifying the preference $P_{\ell} \in \{P_{\ell}^1,P_{\ell}^2,P_{\ell}^3,P_{\ell}^4\}$ for each $\ell \in \{i,j,k\}$ that uniquely induces $\widehat{P}_{\ell}$,
    let $\hat{f}_{\ell}\big(\widehat{P}_i, \widehat{P}_j, \widehat{P}_k\big) = f_{\ell}\big(P_i,P_j,P_k, \widetilde{P}_{-\{i,j,k\}}\big)$ for all $\ell \in \{i,j,k\}$.
    Clearly,
    $\hat{f}$ inherits efficiency, individual rationality and obvious strategy-proofness from $f$.
    
    \medskip

    \noindent
    \textsc{Claim 1}: We have $\hat{f}\big(\widehat{P}_i^4, \widehat{P}_j^1, \widehat{P}_k^1\big)
    = \big\{(i, o_k), (j, o_j), (k, o_i)\big\}$.\medskip
    
    First, note that at the profile $P \equiv (P_i^4, P_j^1, P_k^1, \widetilde{P}_{-\{i,j,k\}})$,
    agent $i$ is the unanimously acclaimed agent, \textit{i.e.},
    $\bar{N}(P) = \{i,j,k\}$,
    $A(P) = \{i,k\}$ and
    $\bar{N}_i(P) = \{j, k\} = \bar{N}(P)\backslash \{i\}$.
    Immediately, by EUA of $f$,
    $\hat{f}_i\big(\widehat{P}_i^4, \widehat{P}_j^1, \widehat{P}_k^1\big)
    = f_i(P)
    = \max\nolimits^{P_i^4}\bar{O}(P) =o_k$.
    This then by individual rationality implies
    $\hat{f}_k\big(\widehat{P}_i^4, \widehat{P}_j^1, \widehat{P}_k^1\big) = o_i$.
    This completes the verification of the claim.\medskip

    \noindent
    \textsc{Claim 2}:
    We have $\hat{f}\big(\widehat{P}_i^4, \widehat{P}_j^1, \widehat{P}_k^2\big) =
    \big\{ (i, o_k), (j, o_j), (k, o_i)\big\}$.\medskip
    
    Similar to Claim 1, by EUA at the preference profile $P \equiv (P_i^4, P_j^1, P_k^2, \widetilde{P}_{-\{i,j,k\}})$,
    we have $\hat{f}_i\big(\widehat{P}_i^4, \widehat{P}_j^1, \widehat{P}_k^2\big)
    = f_i(P) = \max\nolimits^{P_i^4}\bar{O}(P)=  o_k$.
    Next, since $r_1(\widehat{P}_k^2) = o_i$, by Claim 1, strategy-proofness implies $\hat{f}_k\big(\widehat{P}_i^4, \widehat{P}_j^1, \widehat{P}_k^2\big)
    = \hat{f}_k\big(\widehat{P}_i^4, \widehat{P}_j^1, \widehat{P}_k^1\big) = o_i$.
    This completes the verification of the claim.\medskip

    \noindent
    \textsc{Claim 3}:
    We have
    $\hat{f}\big(\widehat{P}_i^3, \widehat{P}_j^2, \widehat{P}_k^2\big)
    =\big\{(i, o_j), (j, o_i), (k, o_k)\big\}$.\medskip
    
    Similar to Claim 1, by EUA at the preference profile $P \equiv (P_i^3, P_j^2, P_k^2, \widetilde{P}_{-\{i,j,k\}})$, we have
    $\hat{f}_i\big(\widehat{P}_i^3, \widehat{P}_j^2, \widehat{P}_k^2\big)
    = f_i(P) = \max\nolimits^{P_i^3}\bar{O}(P) = o_j$.
    Then, individual rationality implies
    $\hat{f}_j\big(\widehat{P}_i^3, \widehat{P}_j^2, \widehat{P}_k^2\big) = o_i$.
    This completes the verification of the claim.\medskip
    
    \noindent
    \textsc{Claim 4}:
    We have
    $\hat{f}\big(\widehat{P}_i^3, \widehat{P}_j^1, \widehat{P}_k^2\big)
    =\big\{(i, o_j), (j, o_i), (k, o_k)\big\}$.\medskip
    
    Similar to Claim 1, by EUA at the preference profile $P \equiv (P_i^3, P_j^1, P_k^2, \widetilde{P}_{-\{i,j,k\}})$, we have
    $\hat{f}_i\big(\widehat{P}_i^3, \widehat{P}_j^1, \widehat{P}_k^2\big)
    = f_i(P) = \max\nolimits^{P_i^3}\bar{O}(P) = o_j$.
    Next, since $r_1(\widehat{P}_j^1) = o_i$, by Claim 3, strategy-proofness implies $\hat{f}_j\big(\widehat{P}_i^3, \widehat{P}_j^1, \widehat{P}_k^2\big)= \hat{f}_j\big(\widehat{P}_i^3, \widehat{P}_j^2, \widehat{P}_k^2\big) = o_i$.
    This completes the verification of the claim.
    \medskip

    \noindent
    \textsc{Claim 5}:
    We have $\hat{f}\big(\widehat{P}_i^3, \widehat{P}_j^4, \widehat{P}_k^3\big)
    =\big\{(i, o_i), (j, o_k), (k, o_j)\big\}$.\medskip
    
    Since $r_1(\widehat{P}_j^4) = r_2(\widehat{P}_k^3)=o_k$ and
    $r_1(\widehat{P}_k^3) = r_2(\widehat{P}_j^4) = o_j$,
    efficiency and individual rationality imply
    $\hat{f}_j\big(\widehat{P}_i^3, \widehat{P}_j^4, \widehat{P}_k^3\big) = o_k$ and $\hat{f}_k\big(\widehat{P}_i^3, \widehat{P}_j^4, \widehat{P}_k^3\big) = o_j$.
    This completes the verification of the claim.
    \medskip
    
    \noindent
    \textsc{Claim 6}: Rule $\hat{f}$ violates obvious strategy-proofness.\medskip
    
    Given $\widehat{\mathcal{D}}_i = \big\{\widehat{P}_i^3, \widehat{P}_i^4\big\}$,
    $\widehat{\mathcal{D}}_j = \big\{\widehat{P}_j^1, \widehat{P}_j^4\big\}$ and
    $\widehat{\mathcal{D}}_k = \big\{\widehat{P}_k^2, \widehat{P}_k^3\big\}$,
    we concentrate on the rule $\hat{f}$ at preference profiles $\big(\widehat{P}_i, \widehat{P}_j, \widehat{P}_k\big) \in \widehat{\mathcal{D}}_i \times \widehat{\mathcal{D}}_j \times \widehat{\mathcal{D}}_k$.
    Since $\hat{f}$ over $\widehat{\mathcal{D}}^3$ is obviously strategy-proof,
    we have an extensive game form $\Gamma$ and a plan $\mathcal{S}_{\ell}: \widehat{\mathcal{D}}_{\ell} \rightarrow S_{\ell}$ for each agent $\ell \in \{i,j,k\}$ that OSP-implement $\hat{f}$ over $\widehat{\mathcal{D}}_i \times \widehat{\mathcal{D}}_j \times \widehat{\mathcal{D}}_k$.
    By the pruning principle,
    we assume w.l.o.g.~that $\Gamma$ is pruned according to
    $\mathcal{S}_i$, $\mathcal{S}_j$ and $\mathcal{S}_k$.
    
    Since $\hat{f}$ is not a constant function,
    by OSP-implementation, $\Gamma$ must have multiple histories.
    Thus, we can assume w.l.o.g.~that at each history, there are at least two actions.
    We focus on the root $h_{\emptyset}$ of $\Gamma$, and let $\rho(h_{\emptyset}) \equiv \ell$.
    There are three cases to consider: $\ell = i$, $\ell = j$ or $\ell = k$.
    Moreover, since $|\widehat{\mathcal{D}}_{\ell}| = 2$ and $|\mathcal{A}(h_{\emptyset})| \geq 2$ in each case,
    by the pruning principle, it must be the case that $|\mathcal{A}(h_{\emptyset})| =2$, and moreover
    the two strategies associated to the two preferences of $\widehat{\mathcal{D}}_{\ell}$ diverge at $h_{\emptyset}$ by choosing the two distinct actions.
    In each case, we induce a contradiction.
    
    First, let $\ell = i$.
    Since we by Claim 5, Claim 2 and OSP-implementation have
    \begin{align*}
        o_i = &~\hat{f}_i\big(\widehat{P}_i^3, \widehat{P}_j^4, \widehat{P}_k^3\big)
        =X_i\big(z^{\Gamma}\big(\mathcal{S}_i^{\widehat{P}_i^3}, \mathcal{S}_j^{\widehat{P}_j^4}, \mathcal{S}_k^{\widehat{P}_k^3}\big)\big)\in X_i\big(h_{\emptyset}, \mathcal{S}_i^{\widehat{P}_i^3}\big)\; \textrm{and}\\
        o_k = &~\hat{f}_i\big(\widehat{P}_i^4, \widehat{P}_j^1, \widehat{P}_k^2\big)
        =X_i\big(z^{\Gamma}\big(\mathcal{S}_i^{\widehat{P}_i^4}, \mathcal{S}_j^{\widehat{P}_j^1}, \mathcal{S}_k^{\widehat{P}_k^2}\big)\big)
        \in X_i\big(h_{\emptyset}, \mathcal{S}_i^{\widehat{P}_i^4}\big),
    \end{align*}
    $o_k \mathrel{\widehat{P}_i^3} o_i$ implies
    $\mathop{\max\nolimits^{\widehat{P}_i^3}}
    X_i\big(h_{\emptyset}, \mathcal{S}_i^{\widehat{P}_i^4}\big)
    \mathrel{\widehat{P}_i^3}
    \mathop{\min\nolimits^{\widehat{P}_i^3}}
    X_i\big(h_{\emptyset}, \mathcal{S}_i^{\widehat{P}_i^3}\big)$ - a contradiction.\medskip

    Second, let $\ell = j$.
    Since we by Claim 2, Claim 5 and OSP-implementation have
    \begin{align*}
        o_j = &~ \hat{f}_j\big(\widehat{P}_i^4, \widehat{P}_j^1, \widehat{P}_k^2\big)
        = X_j\big(z^{\Gamma}\big(\mathcal{S}_i^{\widehat{P}_i^4}, \mathcal{S}_j^{\widehat{P}_j^1}, \mathcal{S}_k^{\widehat{P}_k^2}\big)\big)
        \in X_j\big(h_{\emptyset}, \mathcal{S}_j^{\widehat{P}_j^1}\big)\; \textrm{and}\\
        o_k = &~ \hat{f}_j\big(\widehat{P}_i^3, \widehat{P}_j^4, \widehat{P}_k^3\big)
        =X_j\big(z^{\Gamma}\big(\mathcal{S}_i^{\widehat{P}_i^3}, \mathcal{S}_j^{\widehat{P}_j^4}, \mathcal{S}_k^{\widehat{P}_k^3}\big)\big)
        \in X_j\big(h_{\emptyset}, \mathcal{S}_j^{\widehat{P}_j^4}\big),
    \end{align*}
    $o_k \mathrel{\widehat{P}_j^1} o_j$ implies
    $\mathop{\max\nolimits^{\widehat{P}_j^1}}
    X_j\big(h_{\emptyset}, \mathcal{S}_j^{\widehat{P}_j^4}\big)
    \mathrel{\widehat{P}_j^1}
    \mathop{\min\nolimits^{\widehat{P}_j^1}}
    X_j\big(h_{\emptyset}, \mathcal{S}_j^{\widehat{P}_j^1}\big)$
    - a contradiction.\medskip
    
    Last, let $\ell = k$.
    Since we by Claim 4, Claim 5 and OSP-implementation have
    \begin{align*}
        o_k = &~
        \hat{f}_k\big(\widehat{P}_i^3, \widehat{P}_j^1, \widehat{P}_k^2\big)
        = X_k\big(z^{\Gamma}\big(\mathcal{S}_i^{\widehat{P}_i^3}, \mathcal{S}_j^{\widehat{P}_j^1}, \mathcal{S}_k^{\widehat{P}_k^2}\big)\big)
        \in X_k\big(h_{\emptyset}, \mathcal{S}_k^{\widehat{P}_k^2}\big)\; \textrm{and}\\
        o_j = &~
        \hat{f}_k\big(\widehat{P}_i^3, \widehat{P}_j^4, \widehat{P}_k^3\big)
        =X_k\big(z^{\Gamma}\big(\mathcal{S}_i^{\widehat{P}_i^3}, \mathcal{S}_j^{\widehat{P}_j^4}, \mathcal{S}_k^{\widehat{P}_k^3}\big)\big)
        \in X_k\big(h_{\emptyset}, \mathcal{S}_k^{\widehat{P}_k^3}\big),
    \end{align*}
    $o_j \mathrel{\widehat{P}_k^2} o_k$ implies
    $\mathop{\max\nolimits^{\widehat{P}_k^2}}
    X_k\big(h_{\emptyset}, \mathcal{S}_k^{\widehat{P}_k^3}\big)
    \mathrel{\widehat{P}_k^2}
    \mathop{\min\nolimits^{\widehat{P}_k^2}}
    X_k\big(h_{\emptyset}, \mathcal{S}_k^{\widehat{P}_k^2}\big)$
    - a contradiction.\medskip
    
This completes the verification of the claim and hence proves the lemma.
\end{proof}

\begin{lemma}\label{lem:localSP-extension}
    Given distinct $o_i, o_j, o_k \in O$ such that $o_i \sim o_j$ and $o_j \sim o_k$,
    we have $\overline{(o_i, o_j, o_k)} \in \mathcal{B}(\mathcal{D})$.
\end{lemma}

\begin{proof}
    Since $o_j \sim o_k$, by Lemma \ref{lem:localSP},
    either $\overrightharpoon{(o_i, o_j, o_k)} \in \mathcal{R}(\mathcal{D})$ or
    $\overrightharpoon{(o_i, o_k, o_j)} \in \mathcal{R}(\mathcal{D})$ holds.
    Moreover, since $o_i \sim o_j$, we have a preference $P_{\ell} \in \mathcal{D}$ such that $r_1(P_{\ell}) = o_i$ and $r_2(P_{\ell}) = o_j$, which clearly contradicts the restriction $\overrightharpoon{(o_i, o_k, o_j)}$. Hence, $\overrightharpoon{(o_i, o_j, o_k)} \in \mathcal{R}(\mathcal{D})$.
    Symmetrically, from $o_i \sim o_j$ and $o_k \sim o_j$, we have $\overrightharpoon{(o_k, o_j, o_i)} \in \mathcal{R}(\mathcal{D})$.
    Hence, $\overline{(o_i, o_j, o_k)} \in \mathcal{B}(\mathcal{D})$.
\end{proof}

For the next two lemmas,
let $G$ be an undirected graph over objects where the vertex set is $O$, and two objects form an edge if and only if they are connected.
Given two distinct objects $o, o' \in O$,
a \textit{path} of non-repeated objects $(o^1, \dots, o^q)$ in $G$ connects $o$ and $o'$
if $o^1 = o$, $o^q=o'$, and
$o^k \sim o^{k+1}$ for all $k = 1, \dots, q-1$.
Clearly, by path-connectedness, $G$ is a connected graph, \textit{i.e.}, any two distinct objects are connected by a path in $G$.

\begin{lemma}\label{lem:pathSP}
    Fix a path $(o^1, \dots, o^q)$ in $G$, where $q \geq 3$.
    Given $k \in \{1, \dots, q\}$ and a preference $P_{\ell} \in \mathcal{D}$ such that $r_1(P_{\ell}) = o^k$,
    we have $\big[1< s < k\big] \Rightarrow \big[o^s \mathrel{P_{\ell}} o^{s-1}\big]$ and
    $\big[k< t < q\big] \Rightarrow \big[o^t \mathrel{P_{\ell}} o^{t+1}\big]$.
\end{lemma}

\begin{proof}
    According to the path $(o^1, \dots, o^q)$,
    Lemma \ref{lem:localSP-extension} first implies
    $\overline{(o^{s-1}, o^s, o^{s+1})} \in \mathcal{B}(\mathcal{D})$ for all $s = 2, \dots, q-1$.
    Then, by Lemma \ref{lem:transitivity}, we have
    $\overline{(o^p, o^s, o^t)} \in \mathcal{B}(\mathcal{D})$ for all $1 \leq p < s < t \leq q$.
    Now, given $1< s < k$ and $k< t < q$,
    $\overline{(o^{s-1}, o^s, o^k)} \in \mathcal{B}(\mathcal{D})$ implies $o^s \mathrel{P_{\ell}} o^{s-1}$, and
    $\overline{(o^k, o^t, o^{t+1})} \in \mathcal{B}(\mathcal{D})$ implies $o^t \mathrel{P_{\ell}} o^{t+1}$.
\end{proof}

\begin{lemma}
    Domain $\mathcal{D}$ is a single-peaked domain.
\end{lemma}

\begin{proof}
    The proof consists of two claims.\medskip
    
    \noindent
    \textsc{Claim 1}: The graph $G$ is a tree, \textit{i.e.}, a connected graph that has no cycle.\medskip
    
    Suppose by contradiction that $G$ contains a cycle, \textit{i.e.},
    there exists a path $(o^1, \dots, o^q)$ in $G$ such that $q \geq 3$ and $o^1 \sim o^q$.
    According to the path $(o^1, \dots, o^q)$, Lemma \ref{lem:pathSP} implies $o^2 \mathrel{P_{\ell}}o^q$ for all $P_{\ell} \in \mathcal{D}$ with $r_1(P_{\ell}) = o^1$.
    However, since $o^1 \sim o^q$,
    we have a preference $P_{\ell}' \in \mathcal{D}$ such that $r_1(P_{\ell}') = o^1$ and $r_2(P_{\ell}') = o^q$, which implies $o^q \mathrel{P_{\ell}'} o^2$ - a contradiction.
    This completes the verification of the claim.\medskip
    
    Since $\mathcal{D}$ is a rich domain,
    recall the two completely reversed preferences $\underline{P}_i,\overline{P}_i \in \mathcal{D}$. Let $r_1(\underline{P}_i) \equiv \underline{o}$ and $r_1(\overline{P}_i) \equiv \overline{o}$.
    By Claim 1, we have a unique path $\pi \equiv (o^1, \dots, o^q)$ in $G$ that connects $\underline{o}$ and $\overline{o}$.\medskip
    
    \noindent
    \textsc{Claim 2}: All objects of $O$ are contained in the path $\pi$.\medskip
    
    Suppose not. Thus, since $G$ is a tree by Claim 1, we can identify $o \in O\backslash \{o^1, \dots, o^q\}$ and $k \in \{1, \dots, q\}$ such that $o \sim o^k$.
    There are three cases to consider: (i) $k = 1$, (ii) $k = q$ and (iii) $1< k < q$.
    In each case, we induce a contradiction.
    Note that the first two cases are symmetric. Hence, we focus on cases (i) and (iii).
    In case (i), we have a new path $(o, o^1, \dots, o^q)$ in $G$.
    Immediately, Lemma \ref{lem:pathSP} implies $o^1 \mathrel{P_i} o$ for all $P_i \in \mathcal{D}$ with $r_1(P_i) = o^q = \overline{o}$.
    However, in the preference $\overline{P}_i$, 
    $r_1(\overline{P}_i) = \overline{o}$ and
    $o^1 = \underline{o}$ is bottom-ranked, which implies $o \mathrel{\overline{P}_i} o^1$ - a contradiction.
    In case (iii), we have two new paths $(o^1, \dots, o^k, o)$ and $(o^q, \dots, o^k, o)$.
    Immediately, Lemma \ref{lem:pathSP} implies $o^k \mathrel{P_i} o$ for all $P_i \in \mathcal{D}$ with $r_1(P_i) = o^1 = \underline{o}$, and
    $o^k \mathrel{P_i'} o$ for all $P_i' \in \mathcal{D}$ with $r_1(P_i') = o^q = \overline{o}$.
    However, in the preferences $\underline{P}_i$ and $\overline{P}_i$, we have $r_1(\underline{P}_i) = \underline{o}$, $r_1(\overline{P}_i) = \overline{o}$, and either $o \mathrel{\underline{P}_i} o^k$
    or $o \mathrel{\overline{P}_i} o^k$ - a contradiction.
    This completes the verification of the claim.\medskip

    Clearly, Claim 2 implies that $q = n$. Then, $G$ is a line over $O$ by Claim 1.
    According to the path $\pi$, we construct a linear order $<$ over $O$ such that $o^k < o^{k+1}$ for all $k = 1, \dots, n-1$.
    Immediately, according to the path $\pi$,
    Lemma \ref{lem:pathSP} implies that all preferences of $\mathcal{D}$ are single-peaked w.r.t.~$<$.
    This proves the lemma and hence completes the proof of necessity.
\end{proof}

{\centering\section{OSP-Implementation of Crawler in Example \ref{exm:indispensibility}}\label{app:indispensibility}}

\begin{figure}[h]
    \centering
    \begin{tikzpicture}[scale=0.8,every node/.style={scale=0.8}]
        \node (1) [circle,inner sep=1.5pt,fill=black] at (4,6) {};\node [above of=1,node distance=0.3cm] {$1$};
        \node (2) [circle,inner sep=1.5pt,fill=black] at (0,4.5) {};\node [above left of=2,node distance=0.3cm] {$2$};
        \node (3) [circle,inner sep=1.5pt,fill=black] at (7.2,4.5) {};\node [above right of=3,node distance=0.3cm] {$2$};
        \node (4) [circle,inner sep=1.5pt,fill=black] at (-1,3) {};\node [below of=4,node distance=0.3cm] { $(o_1,o_2,o_3)$};
        \node (5) [circle,inner sep=1.5pt,fill=black] at (0.7,3) {};\node [above right of=5,node distance=0.3cm] {$3$};
        \node (6) [circle,inner sep=1.5pt,fill=black] at (4.2,3) {};\node [above left of=6,node distance=0.3cm] {$1$};
        \node (7) [circle,inner sep=1.5pt,fill=black] at (7.7,3) {};\node [above left of=7,node distance=0.3cm] {$1$};
        \node (8) [circle,inner sep=1.5pt,fill=black] at (12,3) {};\node [above right of=8,node distance=0.3cm] {$3$};
        \node (9) [circle,inner sep=1.5pt,fill=black] at (0,1.5) {};\node [left of=1,below of=9,node distance=0.3cm] { $(o_1, o_3, o_2)$};
        \node (10) [circle,inner sep=1.5pt,fill=black] at (1.5,1.5) {};\node [below of=10,node distance=0.3cm] { $(o_1,o_2,o_3)$};
        \node (11) [circle,inner sep=1.5pt,fill=black] at (3.5,1.5) {};\node [below of=11,node distance=0.3cm] { $(o_2,o_1,o_3)$};
        \node (12) [circle,inner sep=1.5pt,fill=black] at (5,1.5) {};\node [above right of=12,node distance=0.3cm] {$3$};
        \node (13) [circle,inner sep=1.5pt,fill=black] at (7,1.5) {};\node [below of=13,node distance=0.3cm] { $(o_1,o_2,o_3)$};
        \node (14) [circle,inner sep=1.5pt,fill=black] at (8.5,1.5) {};\node [above right of=14,node distance=0.3cm] {$3$};
        \node (15) [circle,inner sep=1.5pt,fill=black] at (10.5,1.5) {};\node [left of=1,below of=15,node distance=0.3cm] { $(o_2,o_3,o_1)$};
        \node (16) [circle,inner sep=1.5pt,fill=black] at (12,1.5) {};\node [below of=16,node distance=0.3cm] { $(o_1,o_3,o_2)$};
        \node (17) [circle,inner sep=1.5pt,fill=black] at (13.5,1.5) {};\node [right of=1, below of=17,node distance=0.3cm] { $(o_1,o_2,o_3)$};
        \node (18) [circle,inner sep=1.5pt,fill=black] at (4.2,-0.5) {};\node [left of=1, below of=18,node distance=0.3cm] { $(o_3,o_1,o_2)$};
        \node (19) [circle,inner sep=1.5pt,fill=black] at (5.8,-0.5) {};\node [below of=19,node distance=0.3cm] { $(o_2,o_1,o_3)$};
        \node (20) [circle,inner sep=1.5pt,fill=black] at (7.7,-0.5) {};\node [below of=20,node distance=0.3cm] { $(o_3,o_2,o_1)$};
        \node (21) [circle,inner sep=1.5pt,fill=black] at (9.3,-0.5) {};\node [right of=2, below of=21,node distance=0.3cm] {$(o_1,o_2,o_3)$};
        \draw[arrow](1) -- node[left=3mm]{$o_1$}(2);
        \draw[arrow](1) -- node[right=2.5mm]{$Pass$}(3);
        \draw[arrow](2) -- node[left=0.5mm]{$o_2$}(4);
        \draw[arrow](2) -- node[right=0.5mm]{$Pass$}(5);
        \draw[arrow](5) -- node[left=0.5mm]{$o_2$}(9);
        \draw[arrow](5) -- node[right=0.5mm]{$o_3$}(10);
        \draw[arrow](3) -- node[left=1.5mm]{$o_1$}(6);
        \draw[arrow](3) -- node[right=0.2mm]{$o_2$}(7);
        \draw[arrow](3) -- node[right=3.5mm]{$Pass$}(8);
        \draw[arrow](6) -- node[left=0.5mm]{$o_2$}(11);
        \draw[arrow](6) -- node[right=0.5mm]{$Pass$}(12);
        \draw[arrow](7) -- node[left=0.5mm]{$o_1$}(13);
        \draw[arrow](7) -- node[right=0.5mm]{$Pass$}(14);
        \draw[arrow](8) -- node[left=0.5mm]{$o_1$}(15);
        \draw[arrow](8) -- node[right=-0.3mm]{$o_2$}(16);
        \draw[arrow](8) -- node[right=0.5mm]{$o_3$}(17);
        \draw[arrow](12) -- node[left=0.5mm]{$o_2$}(18);
        \draw[arrow](12) -- node[right=0.5mm]{$o_3$}(19);
        \draw[arrow](14) -- node[left=0.5mm]{$o_1$}(20);
        \draw[arrow](14) -- node[right=0.5mm]{$o_3$}(21);
    \end{tikzpicture}
    \caption{A millipede game $\Gamma$}\label{fig:GT-eg2}
\end{figure}

We first establish in Figure~\ref{fig:GT-eg2} an extensive game form $\Gamma$ -
a millipede game of \cite{PT2023} which at each decision node has one or more clinch actions and at most one non-clinch action.
Specifically, at each decision node, a clinch action is labeled by an object $o \in O$ which represents that the called agent takes the object $o$ and leaves,
while the non-clinch action (if any) is $Pass$ which means that the playing agent chooses to do nothing but wait for a next move.
At each terminal node, an allocation is attached and specified as a sequence of three objects,
where agent $1$ gets the object at the left, $2$ the middle, and $3$ the right.
Next, each agent's plan is a \textit{greedy-strategy} plan, \emph{that is},
for each $\ell \in N$ and $P_{\ell} \in \mathcal{D}$, at each decision node of agent $\ell$, the strategy $\mathcal{S}_{\ell}^{P_{\ell}}$ chooses the most preferred feasible object if it is a clinch action;
otherwise it chooses the non-clinch action $Pass$.
Then, one can easily verify that the millipede game $\Gamma$ and greedy-strategy plans OSP-implement Crawler $\mathscr{C}: \mathcal{D}^3 \rightarrow \mathcal{M}$ in Example \ref{exm:indispensibility}.

{\centering\section{Proof of Theorem \ref{thm:impossibility}}\label{app:impossibility}}

Suppose by contradiction that there exists a rule $f: \mathcal{D}_<^n \rightarrow \mathcal{M}$, where $n \geq 4$, that satisfies efficiency, individual rationality, obvious strategy-proofness and EBA.
Fixing four agents: $1, 2, 3$ and $4$,
we identify six particular preferences in $\mathcal{D}_<$ and induce six corresponding preferences over $\{o_1, o_2, o_3, o_4\}$,
which are all specified in Table \ref{tab:sixpreferences}.
\begin{table}[h]

    \centering
    \begin{tabular}{ccccccccccccc}
        \hline\hline \rule[0mm]{0mm}{5mm}
        $\overline{P}_{i}^1$ & $\overline{P}_{i}^2$  & $\overline{P}_{i}^3$ & $\overline{P}_{i}^4$ & $\overline{P}_{i}^5$ & $\overline{P}_{i}^6$ &\rule[0mm]{1cm}{0mm}&
        $\widehat{P}_{i}^1$ & $\widehat{P}_{i}^2$  & $\widehat{P}_{i}^3$ & $\widehat{P}_{i}^4$ & $\widehat{P}_{i}^5$ & $\widehat{P}_{i}^6$\\
        \cline{1-6}\cline{8-13}
        $o_1$                  & $o_2$                   & $o_2$                  & $o_3$                  & $o_3$                  & $o_4$ &&
        $o_1$                  & $o_2$                   & $o_2$                  & $o_3$                  & $o_3$                  & $o_4$ \\[-0.2em]
        $o_2$                  & $o_1$                   & $o_3$                  & $o_2$                  & $o_4$                  & $o_3$ &&
        $o_2$                  & $o_1$                   & $o_3$                  & $o_2$                  & $o_4$                  & $o_3$ \\[-0.2em]
        $o_3$                  & $o_3$                   & $o_4$                  & $o_1$                  & $o_2$                  & $o_2$ &&
        $o_3$                  & $o_3$                   & $o_4$                  & $o_1$                  & $o_2$                  & $o_2$ \\[-0.2em]
        $o_4$                  & $o_4$                   & $o_1$                  & $o_4$                  & $o_1$                  & $o_1$ &&
        $o_4$                  & $o_4$                   & $o_1$                  & $o_4$                  & $o_1$                  & $o_1$\\[-0.2em]
        $\vdots$               & $\vdots$                & $\vdots$               & $\vdots$               & $\vdots$               & $\vdots$\\[0.2em]
        \hline\hline
    \end{tabular}
    \caption{Six preferences in $\mathcal{D}_<$ and the induced preferences over $\{o_1, o_2, o_3, o_4\}$}\label{tab:sixpreferences}
\end{table}

\noindent
Let $\overline{\mathcal{D}} = \{\overline{P}_{i}^1, \overline{P}_{i}^2,\overline{P}_{i}^3,\overline{P}_{i}^4,\overline{P}_{i}^5,\overline{P}_{i}^6\}$ and
$\widehat{\mathcal{D}} = \{\widehat{P}_{i}^1, \widehat{P}_{i}^2, \widehat{P}_{i}^3, \widehat{P}_{i}^4, \widehat{P}_{i}^5, \widehat{P}_{i}^6\}$.
For each agent $v\in N\backslash \{1, 2, 3, 4\}$, we fix a preference $\widetilde{P}_{v} \in \mathcal{D}_<$ such that $r_1(\widetilde{P}_{v}) = o_{v}$.
Note that by individual rationality, $f_i\big(\overline{P}_1, \overline{P}_2, \overline{P}_3, \overline{P}_4, \widetilde{P}_{N\backslash \{1, 2,3,4\}}\big) \in \{o_1,o_2,o_3,o_4\}$ for all $i \in \{1,2, 3,4\}$ and $\overline{P}_1, \overline{P}_2, \overline{P}_3, \overline{P}_4 \in \overline{\mathcal{D}}$.
Hence, we can induce a function $\hat{f}$ that allocates objects $o_1, o_2, o_3, o_4$ to agents $1, 2, 3, 4$ according to the induced preferences of $\widehat{\mathcal{D}}$:
for each preference profile $(\widehat{P}_1, \widehat{P}_2, \widehat{P}_3, \widehat{P}_4) \in \widehat{\mathcal{D}}^4$,
after identifying the preference $\overline{P}_{i} \in \overline{\mathcal{D}}$ for each $i \in \{1, 2, 3, 4\}$ that uniquely induces $\widehat{P}_{i}$, let
$\hat{f}_{i}(\widehat{P}_1, \widehat{P}_2, \widehat{P}_3, \widehat{P}_4)
= f_{i}\big(\overline{P}_1, \overline{P}_2, \overline{P}_3, \overline{P}_4, \widetilde{P}_{N\backslash \{1, 2,3,4\}}\big)$
for all $i \in \{1, 2, 3, 4\}$.
Clearly, $\hat{f}$ inherits efficiency, individual rationality and obvious strategy-proofness from $f$.
We first partially characterize $\hat{f}$ via the following two claims.
\medskip

\noindent
\textsc{Claim 1}: We have the following three allocations:
\begin{align*}
    \hat{f}(\widehat{P}_1^6, \widehat{P}_2^6, \widehat{P}_3^1, \widehat{P}_4^1)
    =&~ \big\{(1, o_4), (2,o_3), (3,o_2), (4, o_1)\big\},\\
    \hat{f}(\widehat{P}_1^4, \widehat{P}_2^5, \widehat{P}_3^1, \widehat{P}_4^1)
    =&~ \big\{(1, o_3), (2,o_4), (3,o_1), (4, o_2)\big\}\; \textrm{and}\\
    \hat{f}(\widehat{P}_1^6, \widehat{P}_2^6, \widehat{P}_3^2, \widehat{P}_4^3)
    =&~ \big\{(1, o_3), (2,o_4), (3,o_1), (4, o_2)\big\}.
\end{align*}

Clearly, $\bar{N}(\overline{P}_1^6, \overline{P}_2^6, \overline{P}_3^1, \overline{P}_4^1, \widetilde{P}_{N\backslash \{1, 2,3,4\}})
= \{1, 2, 3, 4\}$ and $A(\overline{P}_1^6, \overline{P}_2^6, \overline{P}_3^1, \overline{P}_4^1, \widetilde{P}_{N\backslash \{1, 2,3,4\}})
= \{1, 4\}$.
By the construction of $\hat{f}$ and EBA of $f$, we have
\begin{align*}
    \hat{f}_1(\widehat{P}_1^6, \widehat{P}_2^6, \widehat{P}_3^1, \widehat{P}_4^1)
    = &~f_1(\overline{P}_1^6, \overline{P}_2^6, \overline{P}_3^1, \overline{P}_4^1, \widetilde{P}_{N\backslash \{1, 2,3,4\}})
    = o_4\; \textrm{and}\\
    \hat{f}_4(\widehat{P}_1^6, \widehat{P}_2^6, \widehat{P}_3^1, \widehat{P}_4^1)
    = &~f_3(\overline{P}_1^6, \overline{P}_2^6, \overline{P}_3^1, \overline{P}_4^1, \widetilde{P}_{N\backslash \{1, 2,3,4\}})
    = o_1.
\end{align*}
Then, efficiency implies $\hat{f}_2(\widehat{P}_1^6, \widehat{P}_2^6, \widehat{P}_3^1, \widehat{P}_4^1) =o_3$ and
$\hat{f}_3(\widehat{P}_1^6, \widehat{P}_2^6, \widehat{P}_3^1, \widehat{P}_4^1) =o_2$.

Symmetrically, we can show $\hat{f}(\widehat{P}_1^4, \widehat{P}_2^5, \widehat{P}_3^1, \widehat{P}_4^1)
= \big\{(1, o_3), (2,o_4), (3,o_1), (4, o_2)\big\}$ and
$\hat{f}(\widehat{P}_1^6, \widehat{P}_2^6, \widehat{P}_3^2, \widehat{P}_4^3)
= \big\{(1, o_3), (2,o_4), (3,o_1), (4, o_2)\big\}$.
This completes the verification of the claim.\medskip

\noindent
\textsc{Claim 2}: We have the following two allocations:
\begin{align*}
    \hat{f}(\widehat{P}_1^6, \widehat{P}_2^5, \widehat{P}_3^2, \widehat{P}_4^3)
    = &~ \big\{(1, o_1), (2,o_3), (3,o_2), (4, o_4)\big\}\; \textrm{and}\\
    \hat{f}(\widehat{P}_1^4, \widehat{P}_2^5, \widehat{P}_3^2, \widehat{P}_4^1)
    = &~ \big\{(1, o_1), (2,o_3), (3,o_2), (4, o_4)\big\}.
\end{align*}

First, 
we know $\hat{f}_2(\widehat{P}_1^6, \widehat{P}_2^4, \widehat{P}_3^3, \widehat{P}_4^3) = o_3$ and
$\hat{f}_3(\widehat{P}_1^6, \widehat{P}_2^4, \widehat{P}_3^3, \widehat{P}_4^3) = o_2$
by individual rationality and efficiency.
Next, since $r_1(\widehat{P}_2^5) = o_3$ and $r_1(\widehat{P}_3^2) = o_2$, by strategy-proofness, we have
$\hat{f}_2(\widehat{P}_1^6, \widehat{P}_2^5, \widehat{P}_3^3, \widehat{P}_4^3) =\hat{f}_2(\widehat{P}_1^6, \widehat{P}_2^4, \widehat{P}_3^3, \widehat{P}_4^3) = o_3$ and
$\hat{f}_3(\widehat{P}_1^6, \widehat{P}_2^4, \widehat{P}_3^2, \widehat{P}_4^3) =\hat{f}_3(\widehat{P}_1^6, \widehat{P}_2^4, \widehat{P}_3^3, \widehat{P}_4^3) = o_2$,
which further by individual rationality imply respectively
$\hat{f}_3(\widehat{P}_1^6, \widehat{P}_2^5, \widehat{P}_3^3, \widehat{P}_4^3) = o_2$ and
$\hat{f}_2(\widehat{P}_1^6, \widehat{P}_2^4, \widehat{P}_3^2, \widehat{P}_4^3) = o_3$.
Immediately, since $r_1(\widehat{P}_3^2) = o_2$ and $r_1(\widehat{P}_2^5) = o_3$, 
$\hat{f}_3(\widehat{P}_1^6, \widehat{P}_2^5, \widehat{P}_3^2, \widehat{P}_4^3)
= \hat{f}_3(\widehat{P}_1^6, \widehat{P}_2^5, \widehat{P}_3^3, \widehat{P}_4^3) = o_2$ and
$\hat{f}_2(\widehat{P}_1^6, \widehat{P}_2^5, \widehat{P}_3^2, \widehat{P}_4^3)
=\hat{f}_2(\widehat{P}_1^6, \widehat{P}_2^4, \widehat{P}_3^2, \widehat{P}_4^3) = o_3$ by strategy-proofness.
Then, we by individual rationality know
$\hat{f}_4(\widehat{P}_1^6, \widehat{P}_2^5, \widehat{P}_3^2, \widehat{P}_4^3)
= o_4$. Thus, we have the allocation $\hat{f}(\widehat{P}_1^6, \widehat{P}_2^5, \widehat{P}_3^2, \widehat{P}_4^3)
= \big\{(1, o_1), (2,o_3), (3,o_2), (4, o_4)\big\}$, as required.
Symmetrically, we can show $\hat{f}(\widehat{P}_1^4, \widehat{P}_2^5, \widehat{P}_3^2, \widehat{P}_4^1)
= \big\{(1, o_1), (2,o_3), (3,o_2), (4, o_4)\big\}$.
This completes the verification of the claim.\medskip

Now, let $\widehat{\mathcal{D}}_1 = \{\widehat{P}_1^4, \widehat{P}_1^6\}$,
$\widehat{\mathcal{D}}_2 = \{\widehat{P}_2^5, \widehat{P}_2^6\}$,
$\widehat{\mathcal{D}}_3 = \{\widehat{P}_3^1, \widehat{P}_3^2\}$ and
$\widehat{\mathcal{D}}_4 = \{\widehat{P}_4^1, \widehat{P}_4^3\}$.
We concentrate on the rule $\hat{f}$ at preference profiles $\big(\widehat{P}_1, \widehat{P}_2, \widehat{P}_3, \widehat{P}_4\big) \in \widehat{\mathcal{D}}_1 \times \widehat{\mathcal{D}}_2 \times \widehat{\mathcal{D}}_3\times \widehat{\mathcal{D}}_4$.
Similar to the proof of Theorem \ref{thm:merit}, 
we have an extensive game form $\Gamma$ and a plan $\mathcal{S}_i: \widehat{\mathcal{D}}_i \rightarrow S_{i}$ for each agent $i \in \{1,2,3,4\}$ that OSP-implement $\hat{f}$ over $\widehat{\mathcal{D}}_1 \times \widehat{\mathcal{D}}_2 \times \widehat{\mathcal{D}}_3\times \widehat{\mathcal{D}}_4$,
where $\Gamma$ is pruned according to $\mathcal{S}_1$, $\mathcal{S}_2$, $\mathcal{S}_3$ and $\mathcal{S}_4$, and there are exactly two actions at the root $h_{\emptyset}$ of $\Gamma$.
Let $\rho(h_{\emptyset}) \equiv i$.
There are four cases to consider: $i = 1$, $i = 2$, $i = 3$ or $i = 4$.
It is clear that the two strategies associated to the two preferences of $\widehat{\mathcal{D}}_i$ diverge at $h_{\emptyset}$ by choosing the two distinct actions.
In each case, we induce a contradiction.

First, let $i = 1$.
Since we by Claim 2, Claim 1 and OSP-implementation know
\begin{align*}
    o_1 = &~\hat{f}_1\big(\widehat{P}_1^6, \widehat{P}_2^5, \widehat{P}_3^2, \widehat{P}_4^3\big)
    =X_1\big(z^{\Gamma}\big(\mathcal{S}_1^{\widehat{P}_1^6}, \mathcal{S}_2^{\widehat{P}_2^5}, \mathcal{S}_3^{\widehat{P}_3^2}, \mathcal{S}_4^{\widehat{P}_4^3}\big)\big)\in X_1\big(h_{\emptyset}, \mathcal{S}_1^{\widehat{P}_1^6}\big)\; \textrm{and}\\
    o_3 = &~\hat{f}_1\big(\widehat{P}_1^4, \widehat{P}_2^5, \widehat{P}_3^1, \widehat{P}_4^1\big)
    =X_1\big(z^{\Gamma}\big(\mathcal{S}_1^{\widehat{P}_1^4}, \mathcal{S}_2^{\widehat{P}_2^5}, \mathcal{S}_3^{\widehat{P}_3^1},\mathcal{S}_4^{\widehat{P}_4^1}\big)\big)
    \in X_1\big(h_{\emptyset}, \mathcal{S}_1^{\widehat{P}_1^4}\big),
\end{align*}
$o_3 \mathrel{\widehat{P}_1^6}o_1$ implies
$\mathop{\max\nolimits^{\widehat{P}_1^6}}
X_1\big(h_{\emptyset}, \mathcal{S}_1^{\widehat{P}_1^4}\big)
\mathrel{\widehat{P}_1^6}
\mathop{\min\nolimits^{\widehat{P}_1^6}}
X_1\big(h_{\emptyset}, \mathcal{S}_1^{\widehat{P}_1^6}\big)$ - a contradiction.\medskip

Second, let $i = 2$.
Since we by Claim 1 and OSP-implementation know
\begin{align*}
    o_4 = &~ \hat{f}_2\big(\widehat{P}_1^4, \widehat{P}_2^5, \widehat{P}_3^1, \widehat{P}_4^1\big)
    = X_2\big(z^{\Gamma}\big(\mathcal{S}_1^{\widehat{P}_1^4}, \mathcal{S}_2^{\widehat{P}_2^5}, \mathcal{S}_3^{\widehat{P}_3^1}, \mathcal{S}_4^{\widehat{P}_4^1}\big)\big)
    \in X_2\big(h_{\emptyset}, \mathcal{S}_2^{\widehat{P}_2^5}\big)\; \textrm{and}\\
    o_3 = &~ \hat{f}_2\big(\widehat{P}_1^6, \widehat{P}_2^6, \widehat{P}_3^1, \widehat{P}_4^1\big)
    =X_2\big(z^{\Gamma}\big(\mathcal{S}_1^{\widehat{P}_1^6}, \mathcal{S}_2^{\widehat{P}_2^6}, \mathcal{S}_3^{\widehat{P}_3^1}, \mathcal{S}_4^{\widehat{P}_4^1}\big)\big)
    \in X_2\big(h_{\emptyset}, \mathcal{S}_2^{\widehat{P}_2^6}\big),
\end{align*}
$o_3 \mathrel{\widehat{P}_2^5} o_4$ implies
$\mathop{\max\nolimits^{\widehat{P}_2^5}}
X_2\big(h_{\emptyset}, \mathcal{S}_2^{\widehat{P}_2^6}\big)
\mathrel{\widehat{P}_2^5}
\mathop{\min\nolimits^{\widehat{P}_2^5}}
X_2\big(h_{\emptyset}, \mathcal{S}_2^{\widehat{P}_2^5}\big)$
- a contradiction.\medskip

Third, let $i = 3$.
Since we by Claim 1 and OSP-implementation have
\begin{align*}
    o_1 = &~ \hat{f}_3\big(\widehat{P}_1^6, \widehat{P}_2^6, \widehat{P}_3^2, \widehat{P}_4^3\big)
    = X_3\big(z^{\Gamma}\big(\mathcal{S}_1^{\widehat{P}_1^6}, \mathcal{S}_2^{\widehat{P}_2^6}, \mathcal{S}_3^{\widehat{P}_3^2}, \mathcal{S}_4^{\widehat{P}_4^3}\big)\big)
    \in X_3\big(h_{\emptyset}, \mathcal{S}_3^{\widehat{P}_3^2}\big)\; \textrm{and}\\
    o_2 = &~ \hat{f}_3\big(\widehat{P}_1^6, \widehat{P}_2^6, \widehat{P}_3^1, \widehat{P}_4^1\big)
    =X_3\big(z^{\Gamma}\big(\mathcal{S}_1^{\widehat{P}_1^6}, \mathcal{S}_2^{\widehat{P}_2^6}, \mathcal{S}_3^{\widehat{P}_3^1}, \mathcal{S}_4^{\widehat{P}_4^1}\big)\big)
    \in X_3\big(h_{\emptyset}, \mathcal{S}_3^{\widehat{P}_3^1}\big),
\end{align*}
$o_2 \mathrel{\widehat{P}_3^2} o_1$ implies
$\mathop{\max\nolimits^{\widehat{P}_3^2}}
X_3\big(h_{\emptyset}, \mathcal{S}_3^{\widehat{P}_3^1}\big)
\mathrel{\widehat{P}_3^2}
\mathop{\min\nolimits^{\widehat{P}_3^2}}
X_3\big(h_{\emptyset}, \mathcal{S}_3^{\widehat{P}_3^2}\big)$
- a contradiction.

Last, let $i = 4$.
Since we by Claim 2, Claim 1 and OSP-implementation have
\begin{align*}
    o_4 = &~\hat{f}_4\big(\widehat{P}_1^4, \widehat{P}_2^5, \widehat{P}_3^2, \widehat{P}_4^1\big)
    = X_4\big(z^{\Gamma}\big(\mathcal{S}_1^{\widehat{P}_1^4}, \mathcal{S}_2^{\widehat{P}_2^5}, \mathcal{S}_3^{\widehat{P}_3^2}, \mathcal{S}_4^{\widehat{P}_4^1}\big)\big)
    \in X_4\big(h_{\emptyset}, \mathcal{S}_4^{\widehat{P}_4^1}\big),\; \textrm{and}\\
    o_2 = &~\hat{f}_4\big(\widehat{P}_1^6, \widehat{P}_2^6, \widehat{P}_3^2, \widehat{P}_4^3\big)
    =X_4\big(z^{\Gamma}\big(\mathcal{S}_1^{\widehat{P}_1^6}, \mathcal{S}_2^{\widehat{P}_2^6}, \mathcal{S}_3^{\widehat{P}_3^2}, \mathcal{S}_4^{\widehat{P}_4^3}\big)\big)
    \in X_4\big(h_{\emptyset}, \mathcal{S}_4^{\widehat{P}_4^3}\big),
\end{align*}
$o_2 \mathrel{\widehat{P}_4^1} o_4$ implies
$\mathop{\max\nolimits^{\widehat{P}_4^1}}
X_4\big(h_{\emptyset}, \mathcal{S}_4^{\widehat{P}_4^3}\big)
\mathrel{\widehat{P}_4^1}
\mathop{\min\nolimits^{\widehat{P}_4^1}}
X_4\big(h_{\emptyset}, \mathcal{S}_4^{\widehat{P}_4^1}\big)$
- a contradiction.
This completes the proof of the theorem.

{\centering\section{Dynamic Individual Rationality}\label{app:DynamicIR}}

\begin{proposition}\label{prop:dynamicIR}
    Given a preference profile $P \in \mathcal{D}_<^n$,
    the Designator allocation $\mathscr{D}(P)$ satisfies ``dynamic individual rationality'', \textit{that is},
    for each $i \in N$,
    \begin{itemize}
    \item[\rm (i)] if $\mathscr{D}_i(P)$ is determined at Step 0 (Preparation), then $\mathscr{D}_i(P)\mathrel{R_i}o_i$;
    
    \item[\rm  (ii)] if $\mathscr{D}_i(P)$ is determined at Step $s \geq 1$,
    then $\mathscr{D}_i(P)\mathrel{R_i}\bar{m}^k(i)$ for all $k = 0, 1, \dots, s-1$.
    \end{itemize}
\end{proposition}

\begin{proof}
We begin with an important observation on the algorithm of Designator.
\medskip

\begin{observation}\label{obs:observation}
Given $P \in \mathcal{D}_<^n$, consider Step $s$ in the algorithm of $\mathscr{D}(P)$, where $1 \leq s < |\bar{N}(P)|$.
Given $i,j \in N_{\bar{m}^{s-1}}$, let the following four conditions hold:
\begin{itemize}
\item[\rm  (i)]
$\max\nolimits^{P_i} O_{\bar{m}^{s-1}} < \bar{m}^{s-1}(i)$, 

\item[\rm (ii)] 
$\bar{m}^{s-1}(\nu)\leq \max\nolimits^{P_{\nu}} O_{\bar{m}^{s-1}} $ for all $\nu \in N_{\bar{m}^{s-1}}$ with 
$\bar{m}^{s-1}(\nu)< \bar{m}^{s-1}(i)$,

\item[\rm (iii)]
$\bar{m}^{s-1}(j)< \bar{m}^{s-1}(i)$, and

\item[\rm (iv)] 
$\bar{m}^{s-1}(\ell)< \max\nolimits^{P_{\ell}} O_{\bar{m}^{s-1}}$ for all $\ell \in N_{\bar{m}^{s-1}}$ with 
$\bar{m}^{s-1}(j) \leq \bar{m}^{s-1}(\ell) < \bar{m}^{s-1}(i)$.
\end{itemize}
Starting from Step $s$, at each step before $i$ leaves, 
an object smaller than $\bar{m}^{s-1}(j)$ is taken and the trivial updating applies.
\end{observation}

    Now, we are ready to show ``dynamic individual rationality'' of Designator.
    By definition, the first condition of dynamic individual rationality is clearly satisfied.
    Next, to show the second condition of dynamic individual rationality, we fix an agent $i \in N$, and let $\mathscr{D}_i(P)$ be determined at Step $s\geq 1$.
    By the algorithm, we know $\mathscr{D}_i(P) = \max\nolimits^{P_i}O_{\bar{m}^{s-1}}$ and $\bar{m}^0(i) \leq \bar{m}^1(i) \leq \dots \leq \bar{m}^{s-1}(i)$.
    If $\bar{m}^0(i) = \bar{m}^1(i) = \dots = \bar{m}^{s-1}(i)$, 
    then $\mathscr{D}_i(P) = \max\nolimits^{P_i}O_{\bar{m}^{s-1}}$ implies $\mathscr{D}_i(P)\mathrel{R_i}\bar{m}^k(i)$ for all $k = 0,1, \dots, s-1$, as required.
    Henceforth, let $\bar{m}^{t-1}(i) < \bar{m}^t(i)$ for some $t \in \{1, \dots, s-1\}$.
    Furthermore, we can assume w.l.o.g.~that $\bar{m}^t(i) = \dots = \bar{m}^{s-1}(i)$.
    Immediately, $\mathscr{D}_i(P) = \max\nolimits^{P_i}O_{\bar{m}^{s-1}}$ implies $\mathscr{D}_i(P) \mathrel{R_i} \bar{m}^k(i)$ for all $k = s-1, \dots, t$.
    In the rest of the proof, we show $\mathscr{D}_i(P) \mathrel{R_i} \bar{m}^k(i)$ for all $k = t-1, \dots, 1,0$.
    
    Since $\bar{m}^{t-1}(i) < \bar{m}^t(i)$, which indicates that in the updating of Step $t$, agent $i$ moves from her current object to a larger object,
    it must be true that $\bar{m}^{t-1}(i) < \bar{m}^{t-1}(i^t) \equiv o^t$, which implies that agent $i$ must find her favorite object in $O_{\bar{m}^{t-1}}$ larger than what she currently holds, \textit{i.e.}, 
    $\bar{m}^{t-1}(i)< \max\nolimits^{P_i}O_{\bar{m}^{t-1}}$. 
    This by single-peakedness of $P_i$ implies $\bar{m}^{t-1}(i)< r_1(P_i)$.
    Thus, we have $\bar{m}^0(i) \leq \bar{m}^1(i) \leq \dots \leq \bar{m}^{t-1}(i) < r_1(P_i)$ which by single-peakedness implies $\bar{m}^{t-1}(i) \mathrel{R_i} \bar{m}^k(i)$ for all $k = t-2, \dots, 1,0$.
    Hence, to complete the verification, by transitivity, it suffices to show
    $\mathscr{D}_i(P) \mathrel{R_i} \bar{m}^{t-1}(i)$.
    
    Clearly, either $\bar{m}^t(i) \mathrel{P_i} \bar{m}^{t-1}(i)$ or $\bar{m}^{t-1}(i) \mathrel{P_i} \bar{m}^t(i)$ holds.
    If $\bar{m}^t(i) \mathrel{P_i} \bar{m}^{t-1}(i)$, then by transitivity, $\mathscr{D}_i(P) \mathrel{R_i} \bar{m}^t(i)$ implies $\mathscr{D}_i(P) \mathrel{R_i} \bar{m}^{t-1}(i)$, as required.
    Henceforth, let $\bar{m}^{t-1}(i) \mathrel{P_i} \bar{m}^t(i)$.
    Recall $\bar{m}^{t-1}(i)< \max\nolimits^{P_i}O_{\bar{m}^{t-1}}$.
    If $\bar{m}^t(i)$ is adjacently larger than $\bar{m}^{t-1}(i)$ in $O_{\bar{m}^{t-1}}$,
    we know $\bar{m}^{t-1}(i)< \bar{m}^t(i) \leq \max\nolimits^{P_i}O_{\bar{m}^{t-1}}$, 
    which by single-peakedness of $P_i$ implies 
    $\bar{m}^t(i)\mathrel{P_i} \bar{m}^{t-1}(i) $ - a contradiction.
    Therefore, $\bar{m}^t(i)$ is never adjacently larger than $\bar{m}^{t-1}(i)$ in $O_{\bar{m}^{t-1}}$, which indicates that at Step $t$, $i$ never crawls from $\bar{m}^{t-1}(i)$ to $\bar{m}^t(i)$, but is designated to inherit $\bar{m}^t(i)$.
    Thus, given $\underline{o}^t \equiv \max\nolimits^{P_{i^t}}O_{\bar{m}^{t-1}} $ and $\bar{o}^t \equiv \bar{m}^{t-1}(i)$, 
    we know $\underline{o}^t \leq \bar{o}^t < o^t = \bar{m}^t(i)$.
    Moreover, since $\bar{o}^t< o^t$ and
    $\bar{o}^t< \max\nolimits^{P_i}O_{\bar{m}^{t-1}}$,
    either $\bar{o}^t< o^t  \leq \max\nolimits^{P_i}O_{\bar{m}^{t-1}}$, or
    $\bar{o}^t< \max\nolimits^{P_i}O_{\bar{m}^{t-1}} < o^t$ holds.
    If $\bar{o}^t< o^t  \leq \max\nolimits^{P_i}O_{\bar{m}^{t-1}}$, single-peakedness of $P_i$ implies $o^t\mathrel{P_i} \bar{o}^t$ (equivalently, $\bar{m}^t(i)\mathrel{P_i} \bar{m}^{t-1}(i)$) - a contradiction.
    Hence, it must be the case that $\bar{o}^t< \max\nolimits^{P_i}O_{\bar{m}^{t-1}} < o^t$.
    Moreover, since $\bar{m}^{t-1}(i^t)$ is not adjacently larger than $\bar{m}^{t-1}(i)$ in $O_{\bar{m}^{t-1}}$, we have an agent $j \in N_{\bar{m}^{t-1}}$ such that $\bar{m}^{t-1}(j)$ is adjacently larger than $\bar{m}^{t-1}(i)=\bar{o}^t$ in $O_{\bar{m}^{t-1}}$. Clearly, 
    $\bar{m}^{t-1}(j)\leq \max\nolimits^{P_i}O_{\bar{m}^{t-1}}$.
    
    Note that if $\mathscr{D}_i(P) = \max\nolimits^{P_i}O_{\bar{m}^{t-1}}$, it immediately implies 
    $\mathscr{D}_i(P) \mathrel{P_i} \bar{m}^{t-1}(i)$.
    Hence, we finally show $\mathscr{D}_i(P) = \max\nolimits^{P_i}O_{\bar{m}^{t-1}}$.
    At Step $t$, after agent $i^t$ takes the object $\underline{o}^t$, the designating updating applies.
    All other agents are fixed to their objects in $O_{\bar{m}^{t-1}}$. Clearly, $\bar{m}^t(i) = o^t$ and $\bar{m}^{t}(j) = \bar{m}^{t-1}(j)$.
    More importantly, we notice the following on Step $t+1$:
    \begin{itemize}
\item[\rm  (i)]
$\max\nolimits^{P_i} O_{\bar{m}^{t}} < \bar{m}^{t}(i)$, 

\item[\rm (ii)] 
$\bar{m}^t(\nu)\leq \max\nolimits^{P_{\nu}} O_{\bar{m}^t} $ for all $\nu \in N_{\bar{m}^t}$ with 
$\bar{m}^t(\nu)< \bar{m}^t(i)$,

\item[\rm (iii)]
$\bar{m}^t(j)< \bar{m}^t(i)$, and

\item[\rm (iv)] 
$\bar{m}^t(\ell)< \max\nolimits^{P_{\ell}} O_{\bar{m}^t}$ for all $\ell \in N_{\bar{m}^t}$ with 
$\bar{m}^t(j) \leq \bar{m}^t(\ell) < \bar{m}^t(i)$.
\end{itemize}
    Immediately, by Observation \ref{obs:observation}, we know that 
    at each step from $t+1$ to $s-1$ (if $s> t+1$), an object smaller than $\bar{m}^{t}(j) = \bar{m}^{t-1}(j)$ is taken, and the trivial updating applies.
    Consequently, at Step $s$ where agent $i$ is called upon to pick her favorite object among the remaining, 
    the object $\max\nolimits^{P_i}O_{\bar{m}^{t-1}}$ remains in $O_{\bar{m}^{s-1}}$, and 
    then agent $i$ takes $\max\nolimits^{P_i}O_{\bar{m}^{t-1}}$ and leaves. Hence, we have $\mathscr{D}_i(P)= \max\nolimits^{P_i}O_{\bar{m}^{t-1}}$, as required.
    This completes the proof.
\end{proof}

{\centering\section{Proof of Theorem \ref{thm:designator}}\label{app:designator}}

It is easy to show that Designator $\mathscr{D}: \mathcal{D}_<^n \rightarrow \mathcal{M}$ satisfies individual rationality and efficiency.
By the proof of the sufficiency part of Theorem \ref{thm:merit}, 
once Designator $\mathscr{D}$ is verified to be obviously strategy-proof, it satisfies EUA,
equivalently, the second condition of WEBA.
In the next lemma, we show that Designator $\mathscr{D}$
satisfies the first condition of WEBA.
\medskip

\begin{lemma}\label{lem:priority}
Given a preference profile $P \in \mathcal{D}_<^n$ with $|A(P)| = 2$, let $A(P) \equiv \{i,j\}$ and $o_i < o_j$.
We have $\mathscr{D}_i(P) = o_j = \max\nolimits^{P_i} \bar{O}(P)$.
\end{lemma}

\begin{proof}
Clearly, $\bar{N}(P)$ is partitioned into the two groups $\bar{N}_i(P)$ and $\bar{N}_j(P)$ and $o_j = \max\nolimits^{P_i}\bar{O}(P)$.
Let $\hat{N} \equiv \{\ell \in \bar{N}_i(P): o_i < o_{\ell}\leq o_j\}$ denote the set of active agents who acclaim $o_i$ and have endowments larger than $o_i$ and weakly smaller than $o_j$.
    Clearly, $j \in \hat{N} $.
    If $\hat{N} = \{j\}$,
    then at Step 1, agent $j$ takes $o_i$ and leaves; agent $i$ is designated to inherit $o_j$.
    Since $o_j = \max\nolimits^{P_i} \bar{O}(P) = \max\nolimits^{P_i} O_{\bar{m}^0}$,
    by dynamic individual rationality of Designator established in Proposition \ref{prop:dynamicIR},
    it must be the case that $\mathscr{D}_i(P) = o_j$, as required. 
Henceforth, let $\hat{N} \supset \{j\}$. 
    We label $\hat{N} = \{\ell_1, \dots, \ell_p\}$, where $p > 1$ and $o_{\ell_k} < o_{\ell_{k+1}}$ for all $k = 1, \dots, p-1$.
    By the definition of Designator, it must be true that all agents of $\hat{N}$ leave in a monotonic order, \textit{that is}, given that agent $\ell_k$ leaves at Step $s_k$ for each $k =1, \dots, p$,
    we have $s_k < s_{k+1}$ for all $k = 1, \dots, p-1$.
     Notice that agent $\ell_1$ takes $o_i$ and leaves at Step 1, which implies $s_1 = 1$, and $i$ is designated to inherit $o_{\ell_1}$.
    Moreover, we notice the following on Step 2:\begin{itemize}
\item[\rm  (i)]
$\max\nolimits^{P_{\ell_2}} O_{\bar{m}^1} < \bar{m}^1(\ell_2)$, 

\item[\rm (ii)] 
$\bar{m}^1(\nu)\leq \max\nolimits^{P_{\nu}} O_{\bar{m}^1} $ for all $\nu \in N_{\bar{m}^1}$ with 
$\bar{m}^1(\nu)< \bar{m}^1(\ell_2)$,

\item[\rm (iii)]
$\bar{m}^1(i)< \bar{m}^1(\ell_2)$, and

\item[\rm (iv)] 
$\bar{m}^1(\ell)< \max\nolimits^{P_{\ell}} O_{\bar{m}^1}$ for all $\ell \in N_{\bar{m}^1}$ with 
$\bar{m}^1(i) \leq \bar{m}^1(\ell) < \bar{m}^1(\ell_2)$.
\end{itemize}
    Immediately, according to Observation \ref{obs:observation} in Appendix \ref{app:DynamicIR},
    it must be true that at each step from 2 to $s_2-1$ (if $s_2-1\geq 2$),
    an object smaller than $\bar{m}^1(i)$ is taken and the trivial updating applies.
    Consequently, at Step $s_2$, $\bar{m}^{s_2-1}(i)= \bar{m}^{s_1}(i) = o_{\ell_1}$, and agent $\ell_2$ takes her favorite object and leaves.
    Furthermore, since $o_i < o_{\ell_1} < o_{\ell_2}$ and $\ell_2 \in \bar{N}_i(P)$ which means $o_i = \max\nolimits^{P_{\ell_2}} \bar{O}(P) = \max\nolimits^{P_{\ell_2}} O_{\bar{m}^0}$, 
    it is true that $r_1(P_{\ell_2})< o_{\ell_1}$ by single-peakedness of $P_{\ell_2}$, which then implies $\max\nolimits^{P_{\ell_2}} O_{\bar{m}^{s_2-1}}\leq o_{\ell_1} = \bar{m}^{s_2-1}(i)$.
    Hence, agent $i$ is recognized as the designated agent at Step $s_2$, and $\bar{m}^{s_2}(i) =o_{\ell_2}$.
    
    By repeatedly applying this argument,
    we realize that for all $k \in \{3, \dots, p\}$, 
    agent $i$ is designated to inherit $o_{\ell_k}$ at Step $s_k$.
    Hence, $\bar{m}^{s_p}(i) = o_{\ell_p} = o_j$.
    Last, since $o_j = \max\nolimits^{P_i} \bar{O}(P) = \max\nolimits^{P_i} O_{\bar{m}^0}$,
    by dynamic individual rationality of Designator,
    it must be the case that $\mathscr{D}_i(P) = o_j$, as required. 
    This proves the Lemma.
\end{proof}

To complete the proof, we now show that Designator $\mathscr{D}: \mathcal{D}_<^n \rightarrow \mathcal{M}$ is obviously strategy-proof.
We first construct an extensive game form $\Gamma$.
\begin{itemize}
    \item[\rm I.] Each decision node is labeled by a 4-tuple $(\kappa, \hat{m}, \bar{m},i)$,
    where (i) $\kappa \in \{\textrm{I}, \textrm{II}\}$,
    (ii) $\hat{m}$ and $\bar{m}$ are sub-allocations such that
    $\bar{m} \neq \emptyset$,
    $\hat{m} \cap \bar{m} = \emptyset$ and
    $\hat{m} \cup \bar{m} \in \mathcal{M}$,
    (iii) $i\in N_{\bar{m}}$, and (iv)
    if $\kappa = \textrm{I}$, then $\hat{m}(\nu) = o_{\nu}$ for all $\nu \in N_{\hat{m}}$ and $\bar{m}(\ell) = o_{\ell}$ for all $\ell \in N_{\bar{m}}$.
    The root $h_{\emptyset}$ is labeled by $(\textrm{I}, \emptyset, e, 1)$.
    It is important to mention that at each decision node $(\kappa, \hat{m}, \bar{m},i)$,
    $\hat{m}$ is the sub-allocation finalized along the previous history initiated from the root $h_{\emptyset}$,
    and $\bar{m}$ is the sub-allocation that needs to be updated according to the subsequent play.
    
    Each terminal node is labeled by an allocation $m \in \mathcal{M}$.
    
    \item[\rm II.]  The player function is $\rho\left(\kappa, \hat{m}, \bar{m}, i\right)=i$ at each decision node $\left(\kappa, \hat{m}, \bar{m}, i\right)$.
    
    \item[\rm III.] At each decision node $(\kappa, \hat{m}, \bar{m},i)$, the action set is specified as follows:
    \begin{itemize}
        \item if $\kappa = \textrm{I}$ and $|N_{\bar{m}}|>1$, then $\mathcal{A}(\kappa, \hat{m}, \bar{m},i) =  \{ \bar{m}(i) \} \cup \{\textrm{Pass}\}$,
        where the action ``$\bar{m}(i)$'' means that agent $i$ leaves with the object $\bar{m}(i)$, while ``Pass'' means that $i$ stays and waits;
        
        \item if $\kappa = \textrm{I}$ and $|N_{\bar{m}}|=1$, then $\mathcal{A}(\kappa, \hat{m}, \bar{m},i) =  \{ \bar{m}(i) \} $;
        
        \item if $\kappa = \textrm{II}$ and $\bar{m}(i)$ is not the largest object in $O_{\bar{m}}$ (\textit{i.e.}, $\bar{m}(i)<\max\nolimits^< O_{\bar{m}}$), then
        an action of $i$ is either ``pick an object $o \in O_{\bar{m}}$ weakly smaller than $\bar{m}(i)$ and leave'', called an \textit{object action}
        or ``pick an object $o\in O_{\bar{m}}$ smaller than $\bar{m}(i)$, leave, and nominate an agent $i^{\ast} \in N_{\bar{m}}$ with $o\leq \bar{m}(i^{\ast})< \bar{m}(i)$ to inherit $\bar{m}(i)$'', called an \textit{object-agent} action, or ``Pass'', \textit{i.e.},
        \begin{align*}
            \mathcal{A}(\kappa, \hat{m}, \bar{m},i) = &~
            \big\{o \in O_{\bar{m}}: o\leq \bar{m}(i)\big\}\\
            &~\cup \big\{ (o, i^{\ast}) \in O_{\bar{m}} \times N_{\bar{m}}: o\leq \bar{m}(i^{\ast})<\bar{m}(i) \big\}
            \cup \{ \textrm{Pass}\};
        \end{align*}

        \item if $\kappa = \textrm{II}$ and $\bar{m}(i)$ is the largest object in $O_{\bar{m}}$ (\textit{i.e.}, $\bar{m}(i)=\max\nolimits^< O_{\bar{m}}$), then
        \begin{align*}
            \mathcal{A}(\kappa, \hat{m}, \bar{m},i) = &~
            \big\{o \in O_{\bar{m}}: o\leq \bar{m}(i)\big\}\\
            &~\cup \big\{ (o, i^{\ast}) \in O_{\bar{m}} \times N_{\bar{m}}: o\leq \bar{m}(i^{\ast})<\bar{m}(i) \big\}.\footnotemark~~~~~~~~~~~~~\,
        \end{align*}
    \end{itemize}
    
    \footnotetext{In particular, if $|N_{\bar{m}} |=1$, then the set of object actions is $\{\bar{m}(i)\}$ and the set of object-agent actions is an empty set. Hence, $\mathcal{A}(\kappa, \hat{m}, \bar{m},i) = \{\bar{m}(i)\}$.}

    \item[\rm IV.] At each decision node $(\textrm{I}, \hat{m}, \bar{m},i)$,
    \begin{itemize}
           \item 
           given that $i$ chooses the action $\bar{m}(i)$, 
           \begin{itemize}
           \item if $|N_{\bar{m}}| = 1$, the game form $\Gamma$ proceeds to the terminal node $m = \hat{m}\cup \big\{\big(i, \bar{m}(i)\big)\big\}$, and
               
           \item if $|N_{\bar{m}}| > 1$, $\Gamma$ proceeds to the decision node $(\textrm{I}, \hat{m}', \bar{m}', j)$ where
            $\hat{m}' = \hat{m}\cup \big\{\big(i, \bar{m}(i)\big)\big\}$,
            $\bar{m}'$ is the trivial updating of $\bar{m}$ (\textit{i.e.}, $\bar{m}' = \bar{m}\backslash \big\{\big(i, \bar{m}(i)\big)\big\}$) and
            $\bar{m}'(j)$ is the smallest object in $O_{\bar{m}'}$;
           \end{itemize}

        \item given that $i$ chooses the action ``Pass'', 
        \begin{itemize}
        \item if $\bar{m}(i)$ is not the largest object in $O_{\bar{m}}$,
        $\Gamma$ proceeds to the decision node $(\textrm{I}, \hat{m}, \bar{m}, j)$ where $\bar{m}(j)$ is adjacently larger than $\bar{m}(i)$ in $O_{\bar{m}}$, and
        
        \item if $\bar{m}(i)$ is the largest object in $O_{\bar{m}}$,
        $\Gamma$ proceeds to the decision node $(\textrm{II}, \hat{m}, \bar{m}, j)$ where
            $\bar{m}(j)$ is the smallest object in $O_{\bar{m}}$.
        \end{itemize}
    \end{itemize}

    \item[\rm V.] At each decision node $(\textrm{II}, \hat{m}, \bar{m},i)$,
\begin{itemize}
\item given that $i$ chooses the action $\bar{m}(i)$, 
\begin{itemize}
\item if $|N_{\bar{m}}| = 1$,
the game form $\Gamma$ proceeds to the terminal node $m = \hat{m}\cup \big\{\big(i, \bar{m}(i)\big)\big\}$, and

\item if $|N_{\bar{m}}| > 1$,
$\Gamma$ proceeds to the decision node $(\textrm{II}, \hat{m}', \bar{m}', j)$ where
            $\hat{m}' = \hat{m}\cup \big\{\big(i, \bar{m}(i)\big)\big\}$,
            $\bar{m}'$ is the trivial updating of $\bar{m}$, and
            $\bar{m}'(j)$ is the smallest object in $O_{\bar{m}'}$;
\end{itemize}
            
\item given that $i$ chooses an object action $o \in O_{\bar{m}}$ with $o< \bar{m}(i)$,
$\Gamma$ proceeds to the decision node $(\textrm{II}, \hat{m}', \bar{m}', j)$ where
            $\hat{m}' = \hat{m}\cup \big\{(i, o)\big\}$, $\bar{m}'(j)$ is the smallest object in $O_{\bar{m}'}$, and
            $\bar{m}'$ is the crawling updating of $\bar{m}$, \textit{i.e.}, 
            \begin{itemize}
            \item each agent of $N_{\bar{m}}$ who has an object in the interval $\big[o, \bar{m}(i)\big)$ crawls to the adjacently larger object in $O_{\bar{m}}$, and
                
            \item all other agents stick to their objects in $O_{\bar{m}}$;
            \end{itemize}

\item given that $i$ chooses an object-agent action $(o, i^{\ast}) \in O_{\bar{m}} \times N_{\bar{m}}$,
where $o \leq \bar{m}(i^{\ast}) < \bar{m}(i)$,
$\Gamma$ proceeds to the decision node $(\textrm{II}, \hat{m}', \bar{m}', j)$ where
            $\hat{m}' = \hat{m}\cup \big\{(i, o)\big\}$,
            $\bar{m}'(j)$ is the smallest object in $O_{\bar{m}'}$, and
            $\bar{m}'$ is the designating updating of $\bar{m}$ w.r.t.~$i^{\ast}$, \textit{i.e.},
            \begin{itemize}
            \item $\bar{m}'(i^{\ast}) = \bar{m}(i)$,
            
            \item each agent of $N_{\bar{m}}$ who has an object in the interval $\big[o, \bar{m}(i^{\ast})\big)$ crawls to the adjacently larger object in $O_{\bar{m}}$, and 
                
            \item all other agents stick to their objects in $O_{\bar{m}}$;
            \end{itemize}
            
\item given that $i$ chooses the action ``Pass'', $\Gamma$ proceeds to the decision node $(\textrm{II}, \hat{m}, \bar{m}, j)$ where
            $\bar{m}(j)$ is adjacently larger than $\bar{m}(i)$ in $O_{\bar{m}}$.

\end{itemize}
\end{itemize}
This completes the construction of $\Gamma$.\medskip

Given a decision node $(\textrm{II}, \hat{m}, \bar{m}, i)$,
along the history from $h_{\emptyset}$ to $(\textrm{II}, \hat{m}, \bar{m}, i)$,
we identify the unique decision nodes $(\textrm{I}, \tilde{m}, \ddot{m}, j)$ and
$(\textrm{II}, \tilde{m}, \ddot{m}, j')$ such that 
$(\textrm{II}, \tilde{m}, \ddot{m}, j')$ is the immediate successor of $(\textrm{I}, \tilde{m}, \ddot{m}, j)$.
Clearly, $\ddot{m}(j)$ is the largest object in $O_{\ddot{m}}$ and 
$\ddot{m}(j')$ is smallest object in $O_{\ddot{m}}$.
Then, we define $O^{\textrm{II}}(\textrm{II}, \hat{m}, \bar{m}, i) = O_{\ddot{m}}$ and
$N^{\textrm{II}}(\textrm{II}, \hat{m}, \bar{m}, i) = N_{\ddot{m}}$.
It is evident that $O^{\textrm{II}}(\textrm{II}, \hat{m}, \bar{m}, i) \supseteq O_{\bar{m}}$ and
$N^{\textrm{II}}(\textrm{II}, \hat{m}, \bar{m}, i) \supseteq N_{\bar{m}}$.\medskip

We next define for each agent $i\in N$ a plan $\mathcal{S}_i: \mathcal{D}_< \rightarrow S_i$.
In this plan, for each preference $P_i \in \mathcal{D}_<$, the strategy $\mathcal{S}_i^{P_i}$ chooses for each decision node $(\kappa, \hat{m}, \bar{m},i)$ an action as follows:
\begin{itemize}
    \item if $\kappa = \textrm{I}$,
    then 
    $\mathcal{S}_i^{P_i}(\kappa, \hat{m}, \bar{m},i) =
    \left\{
    \begin{array}{cl}
    \bar{m}(i)      & \textrm{if}\; \max\nolimits^{P_i} O_{\bar{m}} = \bar{m}(i),\; \textrm{and}\\
    \textrm{Pass} & \textrm{otherwise};
    \end{array}
    \right.$

    \item if $\kappa = \textrm{II}$,
    given $\max\nolimits^{P_i}O^{\textrm{II}}(\kappa, \hat{m}, \bar{m}, i) \equiv o_{i^{\ast}}$ for some $i^{\ast} \in N^{\textrm{II}}(\kappa, \hat{m}, \bar{m}, i)$, then
    \begin{align*}
        \mathcal{S}_i^{P_i}(\kappa, \hat{m}, \bar{m},i) =
        \left\{
        \begin{array}{ll}
        \textrm{Pass} & \textrm{if}\; \bar{m}(i)<\max\nolimits^{P_i}O_{\bar{m}},\\[0.3em]
            \!\big(\max\nolimits^{P_i}O_{\bar{m}}, i^{\ast}\big) & \textrm{if}\; 
            i^{\ast} \in N_{\bar{m}}\; \textrm{and}\; 
            \max\nolimits^{P_i}O_{\bar{m}} \leq \bar{m}(i^{\ast})< \bar{m}(i),~\textrm{and}\\[0.3em]
            \max\nolimits^{P_i}O_{\bar{m}} & \textrm{otherwise}.
        \end{array}
        \right.
    \end{align*}
\end{itemize}
This completes the definition of the plans.
Furthermore, according to the plans $\mathcal{S}_1, \dots, \mathcal{S}_n$,
we prune the extensive game form $\Gamma$.

\begin{lemma}\label{lem:implementation}
    Given a preference profile $P= (P_1, \dots, P_n) \in \mathcal{D}_<^n$,
    let $\mathcal{S}^P \equiv (\mathcal{S}_1^{P_1}, \dots, \mathcal{S}_n^{P_n})$.
    We have $\mathscr{D}(P) = X\big(z^{\Gamma}(\mathcal{S}^{P})\big)$.
\end{lemma}

\begin{proof}
Let $|N\backslash \bar{N}(P)| \equiv \ell$, where $0\leq \ell \leq n$.
    Clearly, by the algorithm of $\mathscr{D}(P)$, we know $\mathscr{D}_i(P) =o_i$ for all $i \in N\backslash \bar{N}(P)$ and the algorithm for $\mathscr{D}_i(P)$ terminates at Step $n-\ell$.
    
   Label $N\backslash \bar{N}(P) =\{\nu_1, \dots, \nu_{\ell}\}$ such that
    $\nu_1 = \min \big\{i \in N: \max\nolimits^{P_i} O = o_i\big\}$, and
    for all $k \in \{2, \dots, \ell\}$,
    $\nu_k = \min \big\{i \in N\backslash \{\nu_1, \dots, \nu_{k-1}\}: \max\nolimits^{P_i} O\backslash \{o_{\nu_1}, \dots, o_{\nu_{k-1}}\} = o_i\big\}$.
    Moreover, denote the terminal history $z^{\Gamma}(\mathcal{S}^{P})$ by
    \begin{align*}
        \big((\textrm{I}, \tilde{m}_1, \ddot{m}_1, 1), \dots, (\textrm{I},\tilde{m}_q, \ddot{m}_q, \eta),
        (\textrm{II}, \hat{m}_1, \bar{m}_1, i_1), \dots, (\textrm{II}, \hat{m}_p, \bar{m}_p, i_p),
        m\big),
    \end{align*}
    where $(\textrm{I}, \tilde{m}_1, \ddot{m}_1, 1) = (\textrm{I}, \emptyset, e, 1)$,
    $m=X\big(z^{\Gamma}(\mathcal{S}^{P})\big)$, $\ddot{m}_n(\eta) = \max\nolimits^< O_{\ddot{m}_q}$ and
    $\bar{m}_1(i_1) = \min\nolimits^< O_{\bar{m}_1}$.
    From the sub-history
    $\big((\textrm{I}, \tilde{m}_1, \ddot{m}_1, 1), \dots, (\textrm{I},\tilde{m}_q, \ddot{m}_q, \eta)\big)$,
    agents $\nu_1, \dots, \nu_{\ell}$ consecutively receive $o_{\nu_1}, \dots, o_{\nu_{\ell}}$.
    Hence, we have $X_{\nu_k}\big(z^{\Gamma}(P)\big) = o_{\nu_k}$ for all $k = 1, \dots, \ell$,
    which implies $\mathscr{D}_{i}(P) = X_{i}\big(z^{\Gamma}(\mathcal{S}^{P})\big)$ for all $i \in N\backslash \bar{N}(P)$ and $\bar{m}_1 = \{(i,o_i): i \in \bar{N}(P)\} = \bar{m}^0$.\footnote{The notation $\bar{m}^0$ is introduced in Step $0$ of the algorithm for $\mathscr{D}(P)$.} 
    Then, we know that for all $k \in \{ 1, \dots, p\}$, 
    $O^{\textrm{II}}(\textrm{II}, \hat{m}_k, \bar{m}_k, i_k)=O_{\bar{m}_1} = O_{\bar{m}^0} = \bar{O}(P)$.
    
    Next, along the sub-history $\big((\textrm{II}, \hat{m}_1, \bar{m}_1, i_1), \dots, (\textrm{II}, \hat{m}_p, \bar{m}_p, i_p)\big)$,
    we identify the smallest index $K_1 \in \{1, \dots, p\}$ such that
    $\mathcal{S}_{i_{K_1}}^{P_{i_{K_1}}}(\textrm{II}, \hat{m}_{K_1}, \bar{m}_{K_1}, i_{K_1}) \neq \textrm{Pass}$.
    From the sub-history
    $\big((\textrm{II}, \hat{m}_1, \bar{m}_1, i_1), \dots,
(\textrm{II}, \hat{m}_{K_1}, \bar{m}_{K_1}, i_{K_1}),
(\textrm{II}, \hat{m}_{K_1+1}, \bar{m}_{K_1+1}, i_{K_1+1})\big)$, we observe
    \begin{itemize}
        \item[\rm (1)]
        $\bar{m}^0 = \bar{m}_1 = \dots = \bar{m}_{K_1}$,
        
        \item[\rm (2)]
        $\bar{m}^0(i_1)< \bar{m}^0(i_2)< \dots < \bar{m}^0(i_{K_1})$,
        
        \item[\rm (3)]
        $\bar{m}^0(i_k) = \bar{m}_k(i_k)<\max\nolimits^{P_{i_k}}O_{\bar{m}_k} = \max\nolimits^{P_{i_k}}O_{\bar{m}^0}$ for all $k = 1, \dots, K_1-1$, 
        
        \item[\rm (4)]
        $\max\nolimits^{P_{i_{K_1}}}O_{\bar{m}^0} =\max\nolimits^{P_{i_{K_1}}}O_{\bar{m}_{K_1}} \leq \bar{m}_{K_1}(i_{K_1}) = \bar{m}^0(i_{K_1})$, which implies $i_{K_1} = i^1$, and
      
      \item[\rm (5)] $X_{i_{K_1}}\big(z^{\Gamma}(\mathcal{S}^P)\big) = \max\nolimits^{P_{i_{K_1}}}O_{\bar{m}_{K_1}} = \max\nolimits^{P_{i^1}} O_{\bar{m}^0}
          =\mathscr{D}_{i^1}(P)$.
      \end{itemize}
      Moreover, let $\max\nolimits^{P_{i_{K_1}}}O_{\bar{m}_{K_1}} = \max\nolimits^{P_{i^1}} O_{\bar{m}^0} \equiv \underline{o}^1$ and $\bar{m}_{K_1}(i_{K_1}) = \bar{m}^0(i^1) \equiv o^1$, and identify $i^{\ast} \in N_{\bar{m}_{K_1}} = N_{\bar{m}^0}$ such that
      $\bar{m}_{K_1}(i^{\ast}) =\bar{m}^0(i^{\ast}) \equiv \underline{o}^1$.
      Clearly, $\underline{o}^1\leq  o^1$. 
      If $\underline{o}^1=  o^1$,
      we have $o_{i^1} = \bar{m}^0(i^1) = o^1 = \underline{o}^1 = \max\nolimits^{P_{i^1}} O_{\bar{m}^0} = \max\nolimits^{P_{i^1}} \bar{O}(P)$ which implies $i^1 \notin \bar{N}(P)$ - a contradiction.
      Hence, we know $\underline{o}^1< o^1$.
      Clearly, $i^{\ast} \in N_{\bar{m}^0}= N^{\textrm{II}}(\textrm{II}, \hat{m}_k, \bar{m}_k, i_k)$
      and $o_{i^{\ast}} =\bar{m}^0(i^{\ast})= \underline{o}^1
      =\max\nolimits^{P_{i_{K_1}}}O_{\bar{m}_{K_1}} 
      =\max\nolimits^{P_{i_{K_1}}}O_{\bar{m}_1}
      = \max\nolimits^{P_{i_{K_1}}} O^{\textrm{II}}(\textrm{II},\hat{m}_{K_1}, \bar{m}_{K_1}, i_{K_1} ) $.
      Then, $i^{\ast} \in N_{\bar{m}_{K_1}}$ and $\max\nolimits^{P_{i_{K_1}}}O_{\bar{m}_{K_1}} =o_{i^{\ast}}< \bar{m}_{K_1}(i_{K_1})$ imply  
      $\mathcal{S}_{i_{K_1}}^{P_{i_{K_1}}}(\textrm{II}, \hat{m}_{K_1}, \bar{m}_{K_1}, i_{K_1}) = \big(\max\nolimits^{P_{i_{K_1}}}O_{\bar{m}_{K_1}}, i^{\ast}) =(\underline{o}^1, i^{\ast})$.
      Hence, $\bar{m}_{K_1+1}$ is the designating updating of $\bar{m}_{K_1}$ w.r.t.~$i^{\ast}$.
      Meanwhile, since $o_{i^{\ast}} = \underline{o}^1 = \max\nolimits^{P_{i^1}} O_{\bar{m}^0} = \max\nolimits^{P_{i^1}} \bar{O}(P)$, we know $i^{\ast} \in A(P)$ and $i^1 \in \bar{N}_{i^{\ast}}(P)$.
      Then, $\underline{o}^1< o^1$, $i^{\ast} \in N_{\bar{m}^0}$ and $\underline{o}^1 =\bar{m}^0(i^{\ast})< \bar{m}_{K_1}(i_{K_1}) =o^1$ (equivalently, $\bar{m}^0(i^{\ast}) \in [\underline{o}^1, o^1)$) imply that $i^{\ast}$ must be recognized as the designated agent at Step 1 of the algorithm for $\mathscr{D}(P)$. 
      Hence, $\bar{m}^1$ is the designating updating of $\bar{m}^0 = \bar{m}_{K_1}$,
      which implies $\bar{m}_{K_1+1}=\bar{m}^1$.

    Analogously,
    we can consecutively identify integers $K_2, \dots, K_{n-\ell} \in \{1, \dots, p\}$,
    where $K_1<K_2< \dots < K_{n-\ell}=p$, such that
    $i_{K_s} = i^s$ and 
    $X_{i_{K_s}}\big(z^{\Gamma}(\mathcal{S}^{P})\big) = \mathscr{D}_{i^s}(P)$ for all $s = 2, \dots, n-\ell$.\footnote{At $(\textrm{II}, \hat{m}_{K_1}, \bar{m}_{K_1}, i_{K_1})$, we infer that $i_{K_1}$ must choose an object-agent action. At $(\textrm{II}, \hat{m}_{K_s}, \bar{m}_{K_s}, i_{K_s})$, where $s \in \{2, \dots, p\}$, agent $i_{K_s}$ may choose an object action, which results in a trivial updating or a crawling updating. Then, it is easy to show that $i_{K_s} = i^s$ and 
    $X_{i_{K_s}}\big(z^{\Gamma}(\mathcal{S}^{P})\big) = \mathscr{D}_{i^s}(P)$.}
    Hence, we have $\mathscr{D}(P) =X\big(z^{\Gamma}(\mathcal{S}^{P})\big)$.
\end{proof}

\begin{lemma}\label{lem:OD}
    Given $i \in N$ and $P_i \in \mathcal{D}_<$\,,
    $\mathcal{S}_i^{P_i}$ is an obviously dominant strategy.
\end{lemma}

\begin{proof}
    Given an arbitrary strategy $s_i \in S_i$,
    we show that $\mathcal{S}_i^{P_i}$ obviously dominates $s_i$ at $P_i$.
    If $\mathcal{E}^{\Gamma}(\mathcal{S}_i^{P_i}, s_i) = \emptyset$,
    $\mathcal{S}_i^{P_i}$ vacuously obviously dominates $s_i$ at $P_i$.
    Henceforth, let $\mathcal{E}^{\Gamma}(\mathcal{S}_i^{P_i}, s_i) \neq \emptyset$.
    We fix an arbitrary decision node $(\kappa, \hat{m}, \bar{m}, i)$ of agent $i$
    such that the history $h_i \equiv \big(h_{\emptyset}, \dots, (\kappa, \hat{m}, \bar{m}, i)\big) \in \mathcal{E}^{\Gamma}(\mathcal{S}_i^{P_i}, s_i)$.
    There are two situations: $\kappa = \textrm{I}$ or $\kappa = \textrm{II}$.
In each situation, we show have $\min\nolimits^{P_i}X_i(h_i, \mathcal{S}_i^{P_i})\mathrel{R_i}\max\nolimits^{P_i}X_i(h_i, s_i)$.
    
    First, let $\kappa = \textrm{I}$.
    Clearly, we have $\mathcal{S}_i^{P_i}(\textrm{I}, \hat{m}, \bar{m}, i) = \bar{m}(i) = o_i$ or
    $\mathcal{S}_i^{P_i}(\textrm{I}, \hat{m}, \bar{m}, i) = \textrm{Pass}$.
    If $\mathcal{S}_i^{P_i}(\textrm{I}, \hat{m}, \bar{m}, i) = \bar{m}(i) = o_i$, 
    we know $\max\nolimits^{P_i} O_{\bar{m}} = o_i$, $X_i(h_i, \mathcal{S}_i^{P_i}) = \{o_i\}$ and $s_i(\textrm{I}, \hat{m}, \bar{m}, i) = \textrm{Pass}$ 
    which implies $X_i(h_i, s_i) \subseteq O_{\bar{m}}$.
    Hence, we have $\min\nolimits^{P_i}X_i(h_i, \mathcal{S}_i^{P_i})\mathrel{R_i}\max\nolimits^{P_i}X_i(h_i, s_i)$, as required.
    Next, let $\mathcal{S}_i^{P_i}(\textrm{I}, \hat{m}, \bar{m}, i) = \textrm{Pass}$.
    Thus, $s_i(\textrm{I}, \hat{m}, \bar{m}, i) = \bar{m}(i) = o_i$ which implies $X_i(h_i, s_i) = \{o_i\}$ and hence $\max\nolimits^{P_i}X_i(h_i, s_i) = o_i$.
    We claim that for all $s_{-i} \in S_{-i}$, $X_i\big(z^{\Gamma}\big(h_i, (\mathcal{S}_i^{P_i},s_{-i})\big)\big) \mathrel{R_i} o_i$ 
    which clearly implies $\min\nolimits^{P_i}X_i(h_i, \mathcal{S}_i^{P_i})\mathrel{R_i}\max\nolimits^{P_i}X_i(h_i, s_i)$.
    Given an arbitrary $s_{-i} \in S_{-i}$,
    let $X\big(z^{\Gamma}\big(h_i, (\mathcal{S}_i^{P_i},s_{-i})\big)\big) \equiv m$ for notational convenience.
    By the pruning principle,
    there exists $(\hat{P}_i, \hat{P}_{-i}) \in \mathcal{D}_<^n$ such that
    the terminal history $z^{\Gamma}(\mathcal{S}_i^{\hat{P}_i},\mathcal{S}_{-i}^{\hat{P}_{-i}}) = \big(h_i, z^{\Gamma}\big(h_i, (\mathcal{S}_i^{P_i},s_{-i})\big)\big)$,
    which implies $X\big(z^{\Gamma}(\mathcal{S}_i^{\hat{P}_i},\mathcal{S}_{-i}^{\hat{P}_{-i}})\big)=m$.
    Since $h_i \in \alpha(\mathcal{S}_i^{P_i})$ by definition,
    $z^{\Gamma}(\mathcal{S}_i^{\hat{P}_i},\mathcal{S}_{-i}^{\hat{P}_{-i}}) = \big(h_i, z^{\Gamma}\big(h_i, (\mathcal{S}_i^{P_i},s_{-i})\big)\big)$ implies 
    $z^{\Gamma}(\mathcal{S}_i^{P_i},\mathcal{S}_{-i}^{\hat{P}_{-i}}) = z^{\Gamma}(\mathcal{S}_i^{\hat{P}_i},\mathcal{S}_{-i}^{\hat{P}_{-i}})$ and hence
    $X\big(z^{\Gamma}(\mathcal{S}_i^{P_i},\mathcal{S}_{-i}^{\hat{P}_{-i}})\big)=m$.
    Immediately, we by Lemma \ref{lem:implementation} have $\mathscr{D}(P_i, \hat{P}_{-i})=m$, which by 
    individual rationality implies $m(i) \mathrel{R_i}o_i$, as required.
    
    Next, let $\kappa = \textrm{II}$.
    Clearly, there are two cases to consider: 
    (i) $\mathcal{S}_i^{P_i}$ chooses an object action or an object-agent action at $(\textrm{II}, \hat{m}, \bar{m}, i)$, or
    (ii) $\mathcal{S}_i^{P_i}(\textrm{II}, \hat{m}, \bar{m}, i) = \textrm{Pass}$.  
    In case (i), 
    by part V of the extensive game form $\Gamma$, we know
    $X_i\big(z^{\Gamma}\big(h_i, (\mathcal{S}_i^{P_i}, s_{-i})\big)\big)
    = \max\nolimits^{P_i}O_{\bar{m}}$ for all $s_{-i} \in S_{-i}$, and hence
    $\min\nolimits^{P_i}X_i(h_i, \mathcal{S}_i^{P_i}) = \max\nolimits^{P_i}O_{\bar{m}}$.
    At $(\textrm{II}, \hat{m}, \bar{m}, i)$, since $X_i\big(z^{\Gamma}\big(h_i, (s_i, s_{-i})\big)\big) \in O_{\bar{m}}$ for all $s_{-i} \in S_{-i}$, it is clear that $\min\nolimits^{P_i}X_i(h_i, \mathcal{S}_i^{P_i})\mathrel{R_i}\max\nolimits^{P_i}X_i(h_i, s_i)$, as required.
    In case (ii),
    we by the construction of $\mathcal{S}_i^{P_i}$ know $\bar{m}(i)< \max\nolimits^{P_i}O_{\bar{m}}$,
    which by single-peakedness implies $\bar{m}(i)<r_1(P_i)$.
    Meanwhile, since at $(\textrm{II}, \hat{m}, \bar{m}, i)$, $s_i$ does not choose the action ``Pass'', 
    we by part V of the extensive game form $\Gamma$ know $X_i(h_i, s_i) = \{o\}$ for some $o \in O_{\bar{m}}$ with $o \leq \bar{m}(i)$.  
    Thus, we have $o \leq \bar{m}(i) < r_1(P_i)$ and hence $\bar{m}(i)\mathrel{R_i}o$ by single-peakedness.
    Therefore, to show $\min\nolimits^{P_i} X_i(h_i, \mathcal{S}_i^{P_i}) \mathrel{R_i} \max\nolimits^{P_i} X_i(h_i, s_i)$, it suffices to show $X_i\big(z^{\Gamma}\big(h_i, (\mathcal{S}_i^{P_i},s_{-i})\big)\big) \mathrel{R_i} \bar{m}(i)$ for all $s_{-i} \in S_{-i}$.
    Given an arbitrary $s_{-i} \in S_{-i}$,
    let $X\big(z^{\Gamma}\big(h_i, (\mathcal{S}_i^{P_i},s_{-i})\big)\big) \equiv m$ for notational convenience.
    By the pruning principle,
    there exists $(\hat{P}_i, \hat{P}_{-i}) \in \mathcal{D}_<^n$ such that
    the terminal history $z^{\Gamma}(\mathcal{S}_i^{\hat{P}_i},\mathcal{S}_{-i}^{\hat{P}_{-i}}) = \big(h_i, z^{\Gamma}\big(h_i, (\mathcal{S}_i^{P_i},s_{-i})\big)\big)$.
    Since $h_i \in \alpha(\mathcal{S}_i^{P_i})$ by definition,
    $z^{\Gamma}(\mathcal{S}_i^{\hat{P}_i},\mathcal{S}_{-i}^{\hat{P}_{-i}}) = \big(h_i, z^{\Gamma}\big(h_i, (\mathcal{S}_i^{P_i},s_{-i})\big)\big)$ implies 
    $z^{\Gamma}(\mathcal{S}_i^{P_i},\mathcal{S}_{-i}^{\hat{P}_{-i}}) = z^{\Gamma}(\mathcal{S}_i^{\hat{P}_i},\mathcal{S}_{-i}^{\hat{P}_{-i}})$ and hence
    $X\big(z^{\Gamma}(\mathcal{S}_i^{P_i},\mathcal{S}_{-i}^{\hat{P}_{-i}})\big)=m$.
    Immediately, we have $\mathscr{D}(P_i, \hat{P}_{-i})=m$ by Lemma \ref{lem:implementation}.
    Note that the decision node $(\textrm{II}, \hat{m}, \bar{m}, i)$ is on the terminal history  $z^{\Gamma}(\mathcal{S}_i^{P_i},\mathcal{S}_{-i}^{\hat{P}_{-i}})$ that implements the Designator allocation $\mathscr{D}(P_i, \hat{P}_{-i})$ by the proof of Lemma \ref{lem:implementation}.
    Then, according to $\mathcal{S}_i^{P_i}(\textrm{II}, \hat{m}, \bar{m}, i) = \textrm{Pass}$ and the algorithm for $\mathscr{D}(P_i, \hat{P}_{-i})$,
    we can identify some Step $t$ with $1\leq t < |\bar{N}(P_i, \hat{P}_{-i})|$ such that $\bar{m}$ equals the sub-allocation $\bar{m}^{t-1}$ which implies $\bar{m}(i) = \bar{m}^{t-1}(i)$, and find that $i$ leaves at some Step $s$ with $s> t$. 
    Hence, we by Proposition \ref{prop:dynamicIR} have $m(i) \mathrel{R_i}\bar{m}(i)$, as required.
    This completes the proof of the lemma and hence of the theorem.
\end{proof}

\bibliographystyle{ecta}
\bibliography{reference} 

\end{document}